\documentclass[preprint,longauthor,longbib,amsmath]{aastex702}
\usepackage[T1]{fontenc}
\usepackage[utf8]{inputenc}
\usepackage{amssymb,booktabs,array,calc}
\usepackage{placeins,needspace}
\usepackage{xurl}
\usepackage{microtype}
\hypersetup{colorlinks=true,linkcolor=xlinkcolor,citecolor=xlinkcolor,urlcolor=xlinkcolor,pdftitle={A Collisional Origin for Ice-rich Iapetus and Titan's Anomalous
Eccentricity},pdfauthor={I. Mosqueira}}
\shorttitle{A collisional origin for Iapetus}
\shortauthors{Mosqueira}
\journalinfo{Draft version September 23, 2026}
\providecommand{\dodoi}[1]{doi:~\href{https://doi.org/#1}{\nolinkurl{#1}}}
\begin{document}
\title{A Collisional Origin for Ice-rich Iapetus and Titan's Anomalous
Eccentricity}
\author{I. Mosqueira}
\affiliation{San Jose State University, One Washington Square, San Jose, CA 95192,
USA}
\email[show]{Ignacio.Mosqueira@sjsu.edu}
\begin{abstract}
We test whether collisions involving Titan can eject enough water-rich
material onto sufficiently distant Saturn-bound orbits to form Iapetus,
while also accounting for Titan's orbital eccentricity. Three-dimensional
impact calculations and post-impact trajectories examine initially
Saturn-unbound and bound impactors. In the oblique
unbound case with a differentiated impactor 
roughly a tenth the mass of Titan at a speed of 10 km s\(^{-1}\),
outgoing predominantly icy material from a selected population
totaling about 3.5 times Iapetus's mass
remains Saturn-bound for days after impact. We track an icy
sample whose orbital apoapsis extends beyond Iapetus's distance.
After impact, Titan's eccentricity increases to near 0.13, 
while the impactor's
rocky core escapes Saturn.
In another case, a bound companion one
quarter Titan's mass collides at \(45^\circ\) and about 3.7 km s\(^{-1}\), 
ejecting
0.60 Iapetus masses of ice. The rocky impactor core initially skips past Titan;
an orbital continuation reaches return contact after 6.9 yr. This run also
retains 0.48 Iapetus masses of ice. Titan's eccentricity evolves due to 
the initial impact and debris scattering. The merger
results in a Titan with an eccentricity of about 0.10. These results
establish the collisional production of an icy debris reservoir, with
some material on orbits of large semimajor axis; 
raising Iapetus's periapse would
require gas drag or dynamical friction.

\end{abstract}

\FloatBarrier
\section{Introduction}\label{introduction}

We explore a collisional origin for Iapetus's ice-rich composition and
Titan's anomalously high orbital eccentricity. In the proposed sequence,
an impact
releases icy material, some of which assembles, scatters outward,
and undergoes
orbital damping. \citet{mosqueira2005} proposed this sequence as
an alternative
to preferential melting, ablation, and delivery of ice-rich Kronian
subnebula disk crossers
(\citealp{mosqueira2010}; see also \citealp{mosqueira2026b}). The two different
scenarios overlap in the observational constraints they address, but are also
complementary in their emphasis. Thus, the ablation scenario
does not address Titan's eccentricity, which would therefore require a
separate explanation, whereas the collisional scenario explored
here considers Iapetus's
composition but does not address the origin of the bulk ice enrichment
of the regular satellites of Jupiter and Saturn compared to the large outer
Solar System bodies (see below). It is therefore possible
that aspects of both scenarios contributed to the observed outcome.

\begin{figure}[!htbp]
\centering
\includegraphics[width=\linewidth,keepaspectratio,alt={Rock-plus-metal mass fraction versus volume-equivalent mean radius, using the same figure as the companion ablation paper (Mosqueira 2026b). The right axis gives the complementary water-rich fraction. Porosity is neglected and compression is included. The comparison supports ice enrichment in Titan, Callisto, and Ganymede as well as the stronger enrichment of Iapetus. Published comparisons (circles): Titan, Sohl et al.~(2003); Ganymede and Callisto, Kuskov and Kronrod (2001, 2005); Pluto and Charon, McKinnon et al.~(2017); Triton, Cioria and Mitri (2022). Compact reference calculations (squares): Mosqueira (2026b), using Fortes (2012) material constants; Table A1 and Appendix A.7 of the companion ablation paper give the values, mass-radius sources, and methods. Vertical bars are the ranges of those particular models or sensitivity calculations, not common statistical error bars. Midpoints are plotting references. The dotted line marks the adopted source's 70\% refractory fraction.}]{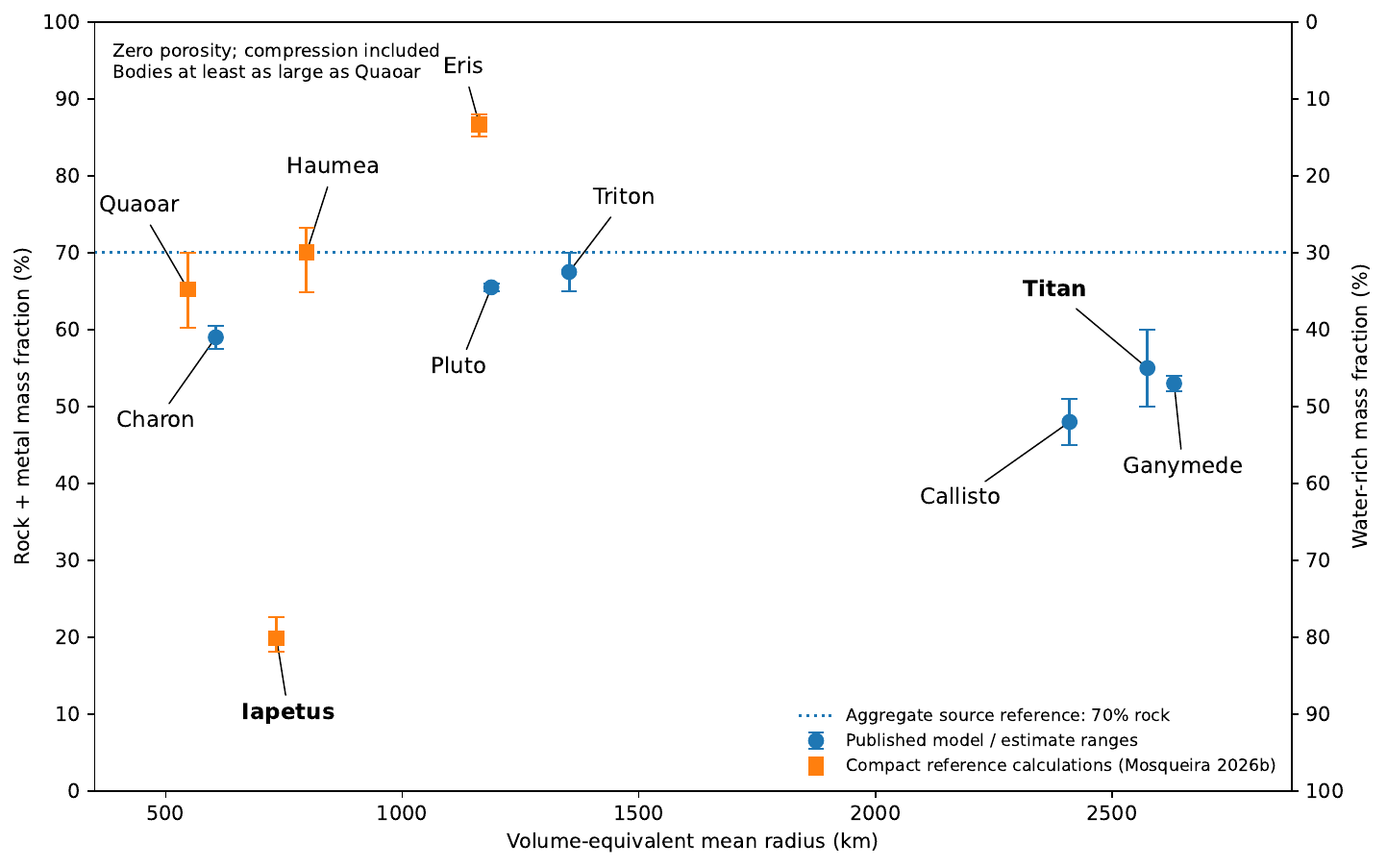}
\caption{Rock-plus-metal mass fraction versus volume-equivalent mean
radius, using the same figure as the companion ablation paper \citep{mosqueira2026b}. The right axis gives the complementary water-rich fraction.
Porosity is neglected and compression is included. The comparison
supports ice enrichment in Titan, Callisto, and Ganymede as well as the
stronger enrichment of Iapetus. Published comparisons (circles): Titan,
\citet{sohl2003}; Ganymede and Callisto, \citet{kuskov2001,kuskov2005}; Pluto and Charon, \citet{mckinnon2017}; Triton, \citet{cioria2022}. Compact reference calculations (squares): \citet{mosqueira2026b}, using \citet{fortes2012} material constants; Table A1 and Appendix
A.7 of the companion ablation paper give the values, mass-radius
sources, and methods. Vertical bars are the ranges of those particular
models or sensitivity calculations, not common statistical error bars.
Midpoints are plotting references. The dotted line marks the adopted
source's 70\% refractory fraction.}\label{fig:observational-context}
\end{figure}

Figure~\ref{fig:observational-context} indicates ice enrichment in Titan, Callisto, and Ganymede
relative to large outer Solar System bodies, including Quaoar, Pluto,
and Triton; Iapetus exhibits a more extreme enrichment.
Quaoar is smaller than Iapetus yet has a much higher bulk density.
Invoking greater porosity in Iapetus would still require explaining its
origin and persistence in the larger body; capture is dynamically
unmotivated and leaves the contrast unexplained.

Hit-and-run collisions are a well-established source of compositional
diversity in planetary accretion \citep{asphaug2006,stewart2012,cambioni2021}.
Tides, shocks, and shear strip mantle material while a core-rich survivor,
or \emph{runner}, escapes \citep{asphaug2006,asphaug2010}. Such stripping
has been proposed for Mercury's large iron core \citep{asphaug2014}.
The analogous ice-rock collision can leave a rock-enriched survivor and
complementary icy debris from either body's mantle. Water-bearing impact
calculations likewise find substantial volatile stripping and
redistribution, with outcomes depending on mass, mass ratio, speed, and
angle \citep{burger2018}.

Runners can return for successive collisions \citep{emsenhuber2019},
including proposed lunar-formation sequences \citep{asphaug2021}.
The central body's gravity can alter or interrupt a grazing pair's
return \citep{emsenhuber2019b}. We calculate the Saturnian return
trajectories independently of terrestrial return probabilities.

Applying this framework requires the mass ratio, contact angle, speed
relative to mutual escape speed, and internal structure. Density
stratification affects collision regimes \citep{gabriel2020}; the
iron-silicate database of \citet{emsenhuber2024} provides an analysis
framework, rather than a calibration of water-rich ejecta yields.

We examine two outcomes for a differentiated impactor: departure of a
refractory survivor, or its retention and possible return to Titan.
The collision and subsequent exchanges with debris and gas together
determine Titan's final orbit and the fate of the icy reservoir.

JPL's SAT441 mean elements give semimajor axes of approximately
\(1.2\times10^6\) km for Titan and \(3.6\times10^6\) km for Iapetus,
with eccentricities 0.029 and 0.028, respectively \citep{jpl2026a}.
We adopt reference locations \(20R_S\) and \(60R_S\), with Saturn's
reference radius \(R_S=6.0\times10^7\) m. The primordial impact radius
and subsequent migration remain scenario variables.

Three-dimensional impacts test compositional partitioning; particle
tracking and gravitational continuations follow refractory escape,
return encounters, and ice retention. Analytical estimates then identify
the assembly, damping, and angular-momentum exchange required for
emplacement.

\FloatBarrier
\section{Physical scenario and conserved
inventories}\label{physical-scenario-and-conserved-inventories}

\subsection{A pre-existing Titan and a differentiated
impactor}\label{a-pre-existing-titan-and-a-differentiated-impactor}

Let \(M_T\) and \(M_I\) denote the present masses of Titan and Iapetus
used as reference scales, \(M_S\) Saturn's mass, and \(G\) the
gravitational constant. Let \(M_0\) and \(m_p\) be the pre-impact
target and projectile masses. The pilot calculations set \(M_0=M_T\)
and \(m_p=0.1M_T\). The reference scales are

\[
M_T\simeq1.3\times10^{23}\ \mathrm{kg},\qquad
M_I\simeq1.8\times10^{21}\ \mathrm{kg},\qquad
GM_S=3.8\times10^{16}\ \mathrm{m^3\,s^{-2}}.
\tag{1}
\]

Main-text results are rounded to two significant figures;
exact counts and numerical settings are retained. Calculations use the
unrounded inputs and saved states documented in the appendices and
accompanying data. Orbital elements \(a,e,i,q,Q\) denote semimajor axis,
eccentricity, inclination to the initial impact plane, pericenter, and
apocenter. Subscripts \(T\) and \(p\) identify Titan (or its primary
remnant) and the projectile.

The satellite masses follow \citet{jpl2026b}. Target and projectile have
rock fractions 0.55 and 0.70, respectively, with distinct rocky cores and
icy mantles. These pilots therefore assume differentiated bodies; the
final remnant mass is an output.

An impactor's source requires a consistent incoming orbit. For
Saturn-centered position \(\boldsymbol r_p\) and velocity
\(\boldsymbol v_p\), its specific two-body energy \(\mathcal E_{p,S}\) is

\[
\mathcal E_{p,S}=\frac12|\boldsymbol v_p|^2-
\frac{GM_S}{|\boldsymbol r_p|}
\tag{2}
\]

with negative and positive values identifying bound and unbound
reference trajectories. The pilots specify encounter geometries rather
than sample their probabilities; their 10 km s\(^{-1}\) speed does not
represent collisions between bound prograde satellites.

\subsection{Component conservation and the useful debris
mass}\label{component-conservation-and-the-useful-debris-mass}

Four immutable tags distinguish target rock, target ice, projectile rock,
and projectile ice. Here \emph{ice} means mantle-origin water; the pilot
equation of state does not determine its post-shock phase.

For each material component \(j\), the initial mass \(M_{j,0}\) is
partitioned at time \(t\) into the dominant remnant (subscript \(T'\)),
other bound aggregates or debris (\(d\)), and material removed from the
satellite-forming region (\(\mathrm{lost}\)):

\[
M_{j,0}=M_{j,T'}(t)+M_{j,d}(t)+M_{j,\mathrm{lost}}(t).
\tag{3}
\]

Outgoing material may remain Saturn bound, reaccrete onto Titan, or join
another aggregate. Its ultimate destination requires following this ledger.

The historical estimate in \citet{mosqueira2005} was an
escaping-ejecta mass \(M_{\rm ej}\) of order \(0.4m_p\). For the reference impactor
this gives

\[
M_{\rm ej}\sim0.04M_T\simeq5.4\times10^{21}\ \mathrm{kg}
\simeq3.0M_I.
\tag{4}
\]

If this ejecta alone supplies Iapetus, the useful fraction
\(\eta_{\rm use}\) ultimately incorporated into its precursor must satisfy

\[
\eta_{\rm use}M_{\rm ej}\geq M_I,\qquad
\eta_{\rm use}\gtrsim0.34.
\tag{5}
\]

These historical scales motivate the calculation. Equation (5) is a
survival requirement after escape, reaccretion, and other losses;
it does not specify a measured collision efficiency.

For initial combined water fraction \(f_0\) and a removed debris mass \(M_d\) with
water fraction \(f_d\), the remnant's water fraction \(f_{T'}\) follows from
conservation:

\[
f_{T'}=\frac{f_0(M_0+m_p)-f_dM_d}{M_0+m_p-M_d},
\tag{6}
\]

when the remaining material forms one dominant remnant. Removing
water-rich debris lowers its water fraction. The combined pre-impact
water fraction in the pilots is 0.44.

\FloatBarrier
\section{Titan's orbital excitation and the impact
calculation}\label{titans-orbital-excitation-and-the-impact-calculation}

\subsection{Impulse scale and orbital diagnostics}\label{an-impulse-scale-not-a-fitted-eccentricity}

Let \(\mu=m_p/M_0\) be the mass ratio and \(\boldsymbol u_{\rm imp}\)
the projectile velocity relative to the target at contact, with magnitude
\(u_{\rm imp}\). An inelastic merger that neglects escaping-ejecta
momentum gives the reference velocity impulse \(\Delta\boldsymbol v_T\).
The circular speed at \(a_T\) is \(v_K\):

\[
\Delta\boldsymbol v_T=\frac{\mu}{1+\mu}\boldsymbol u_{\rm imp},
\qquad v_K=\sqrt{\frac{GM_S}{a_T}}.
\tag{7}
\]

At \(a_T=20R_S\), \(v_K=5.6\) km s\(^{-1}\). With \(\mu=0.1\) and
\(u_{\rm imp}=10\) km s\(^{-1}\), the impulse scale is 0.91 km
s\(^{-1}\). Let \(\Delta v_r,\Delta v_\phi,\Delta v_z\) be its radial,
tangential, and orbit-normal components. For a small impulse applied to
a circular orbit, with \(i_T\) in radians,

\[
e_T^2\simeq
\left(\frac{\Delta v_r}{v_K}\right)^2+
4\left(\frac{\Delta v_\phi}{v_K}\right)^2,
\qquad i_T\simeq\frac{|\Delta v_z|}{v_K}.
\tag{8}
\]

The radial and tangential eccentricity scales are 0.16 and 0.32.
Actual remnants also experience ejecta recoil, incomplete accretion,
and subsequent gravitational exchanges.

For the actual remnant, orbital elements must be calculated from its
Saturn-centered center-of-mass position \(\boldsymbol r_T\) and velocity
\(\boldsymbol v_T\), with \(r_T=|\boldsymbol r_T|\). Define the specific
energy \(\mathcal E_T=|\boldsymbol v_T|^2/2-GM_S/r_T\) and specific
angular momentum \(\boldsymbol h_T=\boldsymbol r_T\times\boldsymbol v_T\).
The eccentricity vector \(\boldsymbol e_T\), of magnitude \(e_T\), and
semimajor axis then follow:

\[
\boldsymbol e_T=
\frac{\boldsymbol v_T\times\boldsymbol h_T}{GM_S}
-\frac{\boldsymbol r_T}{r_T},\qquad
 a_T=-\frac{GM_S}{2\mathcal E_T}.
\tag{9}
\]

Equation (9) supplies the osculating elements reported in Section~\ref{two-collision-pathways}.
Section~\ref{refractory-escape-or-return-and-titans-final-eccentricity} gives the momentum accounting for a return impact.

\subsection{Initial bodies and numerical
method}\label{initial-bodies-and-numerical-method}

The calculations use Spheral's three-dimensional smoothed particle
hydrodynamics (SPH), tree self-gravity, and compatible energy integration
\citep{owen2014,spheral2026}, with a central Saturn potential acting
throughout.

The reference bodies are constructed by an iterative mass-coordinate
hydrostatic calculation at constant initial specific energy
\(2\times10^5\) J kg\(^{-1}\). Antipodal spherical shells sample the
continuous profiles. The resulting target and projectile radii are
\(R_T\simeq2.8\times10^3\) km and \(R_p\simeq1.2\times10^3\) km. The target radius is an outcome of the
adopted composition and equation of state, not an imposed match to
present Titan. The dimensionless impact parameter is \(b=\sin\theta\),
where \(\theta\) is the contact angle measured from head-on.

\begin{table}[!htbp]
\centering
\caption{Inputs to the two impact pilots. Target and projectile use the same
component equations of state. The two geometries are reference cases,
not samples of an inferred impact probability distribution.\label{tab:1}}
\small\setlength{\tabcolsep}{4pt}\renewcommand{\arraystretch}{1.15}
\begin{tabular}{@{}>{\raggedright\arraybackslash}p{(\linewidth - 2\tabcolsep)*\real{0.52000000}}>{\raggedright\arraybackslash}p{(\linewidth - 2\tabcolsep)*\real{0.48000000}}@{}}
\toprule
Input & Reference value\\
\midrule
Target and projectile masses & \(M_T\) and \(0.1M_T\)\\
Target and projectile rock fractions & 0.55 and 0.70\\
Target and projectile particle counts & 2,000 and 200\\
Target and projectile radii & \(2.8\times10^3\) and \(1.2\times10^3\) km\\
Nominal contact speed & 10 km s\(^{-1}\)\\
Impact parameters & \(b=0\) and \(b=0.5\)\\
Initial separation & \(1.3(R_T+R_p)\)\\
Target reference orbit & Circular, \(a_T=20R_S\)\\
Initial rotation of both bodies & Zero\\
Hydrodynamic duration & 3,000 s\\
\bottomrule
\end{tabular}
\end{table}

Native Tillotson basalt and pure-ice models use reference densities of
2,900 and 917 kg m\(^{-3}\), with a zero-pressure floor and no material
strength. Basalt represents the refractory component corresponding to
Figure~\ref{fig:observational-context}'s rock-plus-metal inventory; metallic iron is not modeled
separately. These models follow fluid motion and material identity. The
bounded Gibbs water model does not cover the full shock-release domain
and is not used.

We use a cubic spline kernel, two particles per smoothing length,
artificial-viscosity coefficients 1 and 2, compatible energy evolution,
a second-order Runge--Kutta integrator, and Courant factor 0.2. The
completed pilots use density summation. A separate density-integration
attempt is described in Appendix~\ref{appendix-a.-numerical-controls-and-interpretation-of-the-pilot-results}. The discrete initial bodies were not
independently relaxed; isolated controls therefore accompany the impact
runs and delimit their present quantitative interpretation.

\subsection{Geometry and velocity
conventions}\label{geometry-and-velocity-conventions}

The impact parameter is related to the relative specific angular
momentum \(h_{\rm rel}\) by

\[
b=\frac{h_{\rm rel}}{(R_T+R_p)u_{\rm imp}}.
\tag{10}
\]

The initial relative speed \(u_0\) at separation \(d_0=1.3(R_T+R_p)\) follows the
two-body contact-energy relation,

\[
u_0^2=u_{\rm imp}^2-
2G(M_0+m_p)\left[\frac{1}{R_T+R_p}-\frac{1}{d_0}\right].
\tag{11}
\]

\begin{figure}[!htbp]
\centering
\includegraphics[width=\linewidth,keepaspectratio,alt={Oblique Spheral impact at 0, 1,500, and 3,000 s. All particles are projected onto the impact plane; colors distinguish original body and component. The origin follows the target-material center of mass. Markers are numerical particles, not measured physical fragment sizes.}]{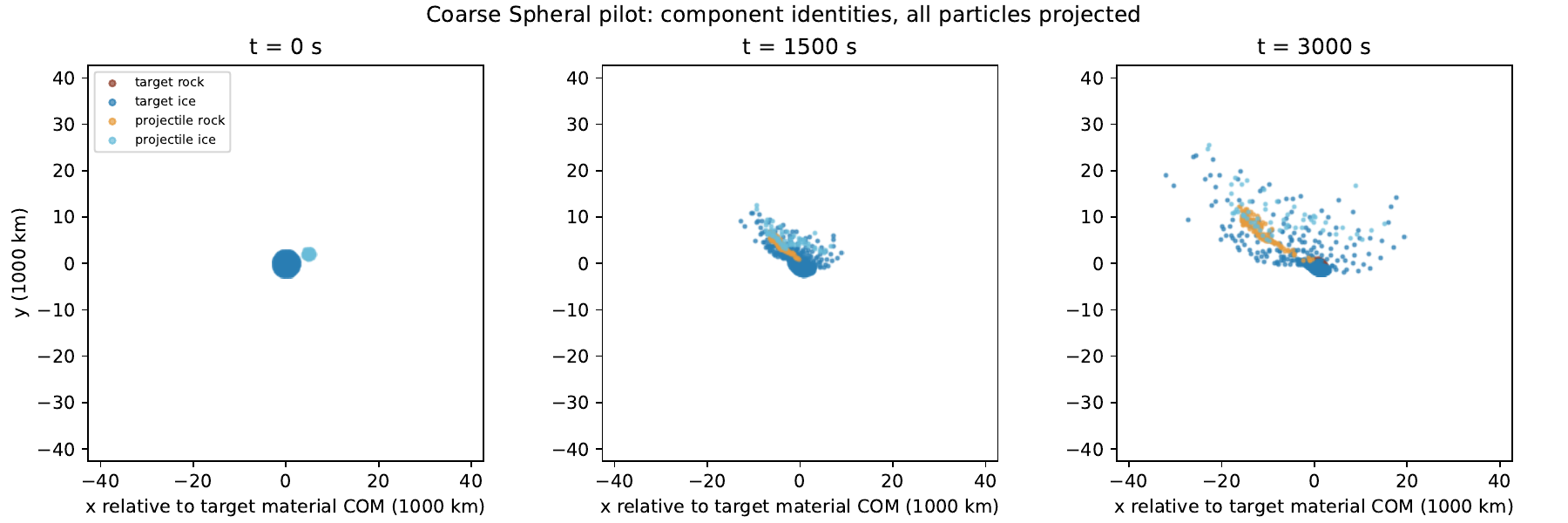}
\caption{Oblique Spheral impact at 0, 1,500, and 3,000 s. All particles
are projected onto the impact plane; colors distinguish original body
and component. The origin follows the target-material center of mass.
Markers are numerical particles, not measured physical fragment
sizes.}\label{fig:impact}
\end{figure}

The approach is radial in Saturn's orbital plane, with a prograde
transverse offset in the oblique case and no initial spin. The mutual
escape speed \(v_{\rm esc}\simeq2.2\) km s\(^{-1}\) is well below the
contact speed. The projectile center of mass (COM) has specific
Saturn-centered energy approximately \(3.4\times10^7\) J kg\(^{-1}\)
in both pilots, identifying an unbound incoming encounter.
Figure~\ref{fig:impact} shows the oblique impact.

\subsection{A glancing collision with a bound
satellite}\label{a-glancing-collision-with-a-bound-satellite}

For the bound branch we seek a glancing collision that releases ice,
followed by return and incorporation of the refractory survivor into
Titan. We use companion-to-target mass ratios 0.10 and 0.25, the
compositions and material models of Table~\ref{tab:1}, and 2,000 target particles
with 200 or 500 companion particles. The larger companion has radius
\(1.7\times10^3\) km and \(5.6M_I\) of ice; the smaller has
\(2.2M_I\). Satellite-impact calculations also exhibit merging,
hit-and-run, and fragmentation regimes \citep{sekine2012}, with yields
dependent on composition and mass ratio.

Titan starts on a circular orbit at \(20R_S\). The coplanar prograde
companion has \(q_p=10R_S\), \(Q_p=30R_S\), \(a_p=20R_S\), and
\(e_p=0.5\), representing intersecting orbits after excitation. From
approximately 3.1 mutual Hill radii of separation, we integrate both
centers under mutual and Saturnian gravity. Adjusting the companion's
longitude selects \(b=0.87\), or \(60^\circ\) from head-on; contact
speeds are about 3.7 km s\(^{-1}\) for both masses. Both bodies
remain Saturn bound through the hydrodynamic handoff. Speed follows from
the orbits and mutual focusing. A \(45^\circ\) case uses the larger
companion and the same initial orbital elements.

Hydrodynamics starts at \(1.3(R_T+R_p)\), including Titan's approach
recoil. The preceding center integration uses the same softened gravity
without deformation. The orbital handoff collapses the identified
primary and mechanically self-bound survivor to their centers, retains
every other numerical sample, and preserves component masses and COM
moments while archiving internal energy and spin. These pressure-free
continuations use IAS15 in REBOUND 5.1.1 \citep{rein2015}. They omit gas, solar
perturbations, cooling, and higher Saturnian harmonics, but include the
fixed-axis quadrupole coefficient \(J_2\simeq0.016\) at reference radius
\(R_{\rm ref}\simeq6.0\times10^4\) km \citep{iess2019}; using the
present coefficient is an approximation for the early system.

An approach screen supplies geometry for a subsequent collision
calculation. Extended \(45^\circ\) continuations use compact radii from
component volumes and integrate returning pairs to first surface
contact, stopping before hydrodynamic merger. Appendix~\ref{appendix-c.-numerical-debris-controls-and-physical-fragment-assumptions} specifies the
force and event definitions.

\FloatBarrier
\section{Two collisional pathways}\label{two-collision-pathways}

\subsection{Initially unbound impacts: ice release and material
origins}\label{initially-unbound-impacts-ice-release-and-material-origins}

At each output we identify an iteratively mechanically bound dominant
remnant. For particle \(k\), let \(r_{k,T'}\), \(\dot r_{k,T'}\), and
\(\mathcal E_{k,T'}\) be its remnant-relative distance, radial velocity,
and specific mechanical energy. Let \(R_{95,T'}\) be the remnant's
95th-percentile particle radius and \(\mathcal E_{k,S}\) the particle's
specific Saturn-centered two-body energy. The outgoing selection requires:

\[
\dot r_{k,T'}>0,\quad \mathcal E_{k,T'}>0,\quad
r_{k,T'}>2R_{95,T'},\quad \mathcal E_{k,S}<0.
\tag{12}
\]

This instantaneous selection excludes internal energy and later pressure
work. It distinguishes escape from Titan from escape from Saturn;
subsequent reaccretion or other exchanges remain possible.

\begin{table}[!htbp]
\centering
\caption{Selected material at 3,000 s in the two 10 km s\(^{-1}\) pilots. Ice
fraction refers to immutable material tags. The last two columns
respectively give Titan's contribution to the selected ice and the
fraction of the original projectile rock incorporated into the dominant
remnant. Masses and percentages are rounded independently from full
precision.\label{tab:2}}
\small\setlength{\tabcolsep}{4pt}\renewcommand{\arraystretch}{1.15}
\begin{tabular}{@{}>{\raggedright\arraybackslash}p{(\linewidth - 8\tabcolsep)*\real{0.20000000}}>{\raggedright\arraybackslash}p{(\linewidth - 8\tabcolsep)*\real{0.20000000}}>{\raggedright\arraybackslash}p{(\linewidth - 8\tabcolsep)*\real{0.20000000}}>{\raggedright\arraybackslash}p{(\linewidth - 8\tabcolsep)*\real{0.20000000}}>{\raggedright\arraybackslash}p{(\linewidth - 8\tabcolsep)*\real{0.20000000}}@{}}
\toprule
Geometry & Selected mass (\(M_I\)) & Ice fraction & Titan share of ice & Projectile rock retained\\
\midrule
Head-on & 2.1 & 1 & 79\% & 93\%\\
Oblique, \(b=0.5\) & 3.6 & 0.80 & 95\% & 3.6\%\\
\bottomrule
\end{tabular}
\end{table}

\begin{figure}[!htbp]
\centering
\includegraphics[width=0.83\linewidth,keepaspectratio,alt={Component masses in the selected Saturn-bound, outward-moving population at 3,000 s. The bars use the four-component particle tally underlying Table 2: the head-on case has 2.1 Iapetus masses of ice-origin material and no selected rock-origin material, while the oblique case has approximately 2.9 Iapetus masses and 0.71 Iapetus masses, respectively. The one-Iapetus-mass line is a scale comparison, not an accreted-mass result. Projectile core retention is reported separately in Table 2.}]{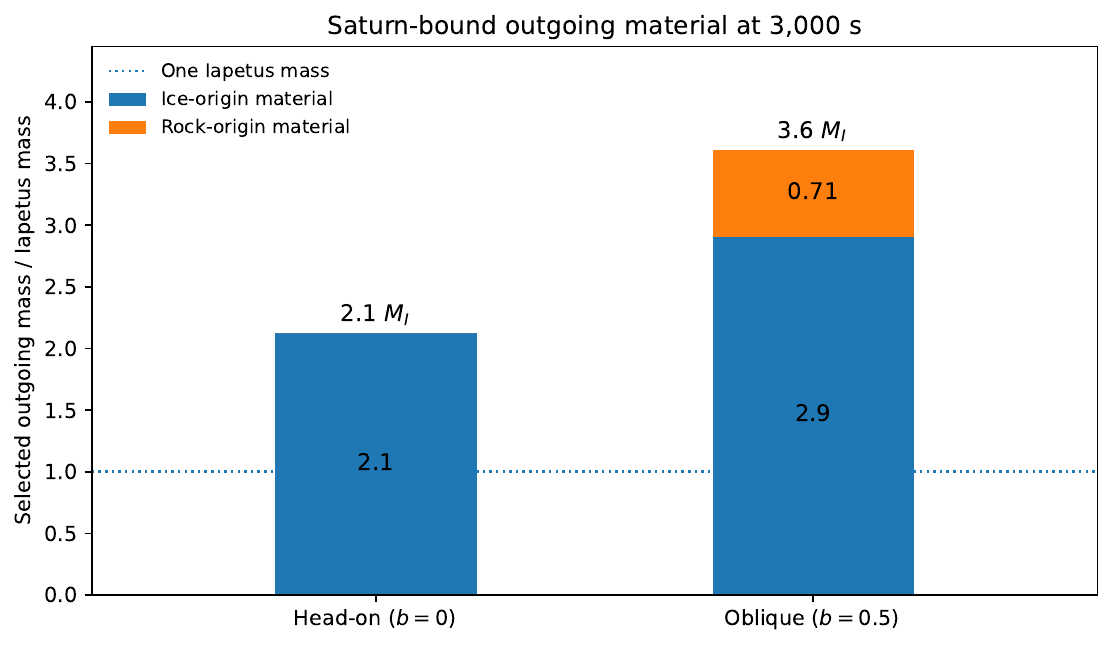}
\caption{Component masses in the selected Saturn-bound, outward-moving
population at 3,000 s. The bars use the four-component particle tally
underlying Table~\ref{tab:2}: the head-on case has \(2.1M_I\) of ice-origin
material and no selected rock-origin material, while the oblique case
has approximately \(2.9M_I\) and \(0.71M_I\), respectively. The
one-Iapetus-mass line is a scale comparison, not an accreted-mass
result. Projectile core retention is reported separately in Table~\ref{tab:2}.}\label{fig:partition}
\end{figure}

Titan's mantle supplies most selected ice (Table~\ref{tab:2}; Figure~\ref{fig:partition}). The
head-on sample contains \(1.7M_I\) of target ice and \(0.45M_I\) of
projectile ice; the oblique sample contains \(2.8M_I\), \(0.15M_I\),
and \(0.71M_I\) of projectile rock. Neither selection contains target
rock. At 3,000 s, the dominant remnants have
\((a_T/R_S,e_T)=(21,0.20)\) and \((18,0.13)\), respectively.

The oblique projectile's complete rock tally closes: of its 140 original
elements, five belong to the dominant remnant, 19 enter the selected
outgoing subset, three are otherwise Saturn bound outside the remnant,
and 113 have positive Saturn-centered energy outside it. These groups
contain \(0.19\), \(0.71\), \(0.11\), and \(4.2M_I\),
respectively. Section~\ref{refractory-departure-the-complete-inventory-and-titans-excitation} follows the departing refractory material.

The oblique geometry has nominal \(\theta=30^\circ\) and mass ratio 0.1,
matching one example in \citet[Figure~3b]{asphaug2006},
but differs in speed and materials. For our stated masses and radii,

\[
v_{\rm esc}=\left[\frac{2G(M_0+m_p)}{R_T+R_p}\right]^{1/2}
=2.2\ \mathrm{km\,s^{-1}},\qquad u_{\rm imp}/v_{\rm esc}=4.5.
\tag{12a}
\]

With \(M_\oplus\) denoting Earth's mass, their iron-silicate example
uses a \(0.1M_\oplus\) target and twice the mutual escape speed. Its
yields cannot be transferred to these faster ice-rock impacts.

Material provenance provides a further distinction. In the examples of
\citet{asphaug2006}, both bodies lose mantle, with particular emphasis
on stripping and disruption of the smaller impactor. Target-derived
orbiting debris nevertheless has precedent: \citet{reufer2012} obtain
protolunar disks enriched in excavated target mantle while much of the
impactor escapes, as also discussed by \citet{asphaug2021}. In our
unbound-impact pilots, Titan supplies 95\% of the selected ice in the
oblique case and 79\% in the head-on case. These fractions describe the
Saturn-bound outgoing ice selected by equation (12), not all material
removed from both bodies. The established hit-and-run framework explains
compositional separation; Titan's dominant contribution to this icy
reservoir is an outcome of the particular impacts and orbital selection
examined here.

Our numerical elements each carry \(0.037M_I\), only about 27 elements per Iapetus
mass, so the pilots do not resolve a fragment spectrum or exact purity.
Appendix~\ref{appendix-a.-numerical-controls-and-interpretation-of-the-pilot-results} gives the numerical controls.

\subsection{Refractory departure, the complete inventory, and
Titan's
excitation}\label{refractory-departure-the-complete-inventory-and-titans-excitation}

Fixed indices identify nested 16- and 43-element projectile-rock cohorts
that remain mechanically self-bound from 1,000 to 3,300 s. Their masses
are \(0.60M_I\) and \(1.6M_I\); both describe the same concentration.
At 3,300 s their COM Saturn energies are +0.89 and +1.7 MJ kg\(^{-1}\).
Table~\ref{tab:C1} and Appendix~\ref{appendix-c.-numerical-debris-controls-and-physical-fragment-assumptions} give internal structure and grouping controls.

The five-day reciprocal-gravity continuation collapses only the compact
1,808-element primary; all other 392 samples gravitate individually.
The primary's 1,100 target-rock, 704 target-ice, and four projectile-rock
elements give mass \(M_{T'}=0.90M_T=1.2\times10^{23}\) kg and rock
fraction 0.61. Its membership stays fixed, and the rock cohorts remain
tracking sets. This calculation omits pressure, collisions, gas, and
solar perturbations (Table~\ref{tab:A1}).

\begin{figure}[!htbp]
\centering
\includegraphics[width=\linewidth,keepaspectratio,alt={Conditional five-day continuation with reciprocal gravity, starting at the 3,000-s impact output. Left: Saturn-centered COM energies of the fixed nested rock cohorts; fluctuations include internal exchanges, and neither cohort is a prescribed rigid object. Right: negative-energy mass from the original 97-element selection, its ice-tagged part, and the subset without a recorded Saturn or primary-radius crossing. All 392 unresolved samples gravitate; pressure, collisions, and gas are omitted. The tighter integration agrees on the final selected-material membership and outbound positive-energy state, but not detailed internal trajectories.}]{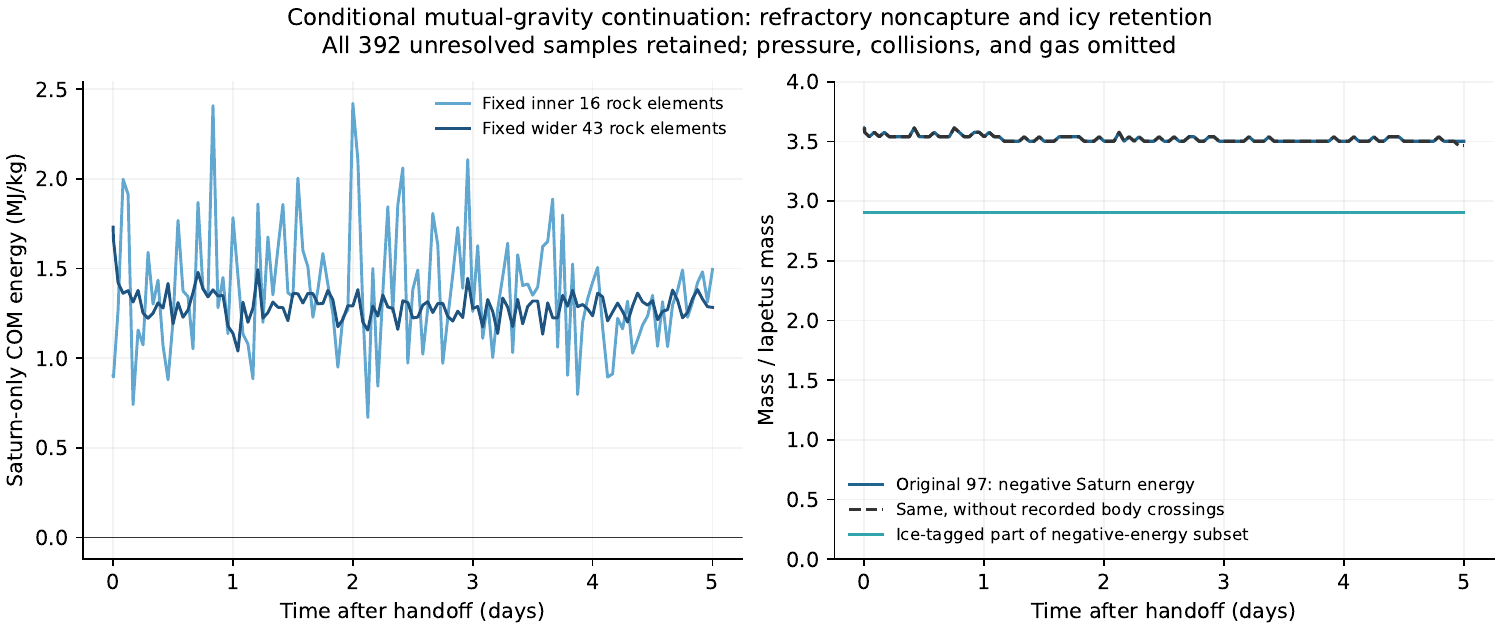}
\caption{Conditional five-day continuation with reciprocal gravity,
starting at the 3,000-s impact output. Left: Saturn-centered COM
energies of the fixed nested rock cohorts; fluctuations include internal
exchanges, and neither cohort is a prescribed rigid object. Right:
negative-energy mass from the original 97-element selection, its
ice-tagged part, and the subset without a recorded Saturn or
primary-radius crossing. All 392 unresolved samples gravitate; pressure,
collisions, and gas are omitted. The tighter integration agrees on the
final selected-material membership and outbound positive-energy state,
but not detailed internal trajectories.}\label{fig:escape-retention}
\end{figure}

The rock passes Saturn pericenter about 1.5 days after handoff and
reaches about \(34R_S\) outbound at five days (Figure~\ref{fig:escape-retention}). The wider
cohort has Saturn energy +1.3 MJ kg\(^{-1}\) and radial velocity
+4.9 km s\(^{-1}\); the tighter integration gives +1.3 MJ kg\(^{-1}\)
and +4.9 km s\(^{-1}\). Neither cohort records a Saturn or primary
crossing. The core is being lost while ice remains bound: refractory
noncapture, since the projectile was already Saturn unbound.

Of the original 97 selected elements, the same 94 retain negative Saturn
energy in both integrations: \(3.5M_I\), with ice fraction 0.83.
All 78 selected ice elements remain negative-energy; the three sign
changes are projectile-rock members of the wider cohort. One selected
ice element records a primary-radius crossing. Excluding it leaves
\(3.5M_I\) with ice fraction 0.83. These numbers follow the
original selection, rather than all later negative-energy material.

\begin{table}[!htbp]
\centering
\caption{Complete inventory of the 392 unresolved elements at five days in the
reference reciprocal-gravity run. Entries in the three material columns
are element counts; all elements have mass \(0.037M_I\). There is no
target rock outside the compact primary. Rows are disjoint and include
every unresolved element. A crossing is a geometric flag, not accretion
or removal. Masses are rounded independently from full precision.\label{tab:3}}
\small\setlength{\tabcolsep}{4pt}\renewcommand{\arraystretch}{1.15}
\begin{tabular}{@{}>{\raggedright\arraybackslash}p{(\linewidth - 8\tabcolsep)*\real{0.34000000}}>{\raggedright\arraybackslash}p{(\linewidth - 8\tabcolsep)*\real{0.14000000}}>{\raggedright\arraybackslash}p{(\linewidth - 8\tabcolsep)*\real{0.18000000}}>{\raggedright\arraybackslash}p{(\linewidth - 8\tabcolsep)*\real{0.17000000}}>{\raggedright\arraybackslash}p{(\linewidth - 8\tabcolsep)*\real{0.17000000}}@{}}
\toprule
Saturn energy / recorded crossing & Titan ice & Projectile rock & Projectile ice & Total mass (\(M_I\))\\
\midrule
Negative, no crossing & 87 & 24 & 6 & 4.4\\
Negative, primary crossing & 27 & 1 & 0 & 1.0\\
Nonnegative, no crossing & 82 & 111 & 54 & 9.2\\
Nonnegative, any crossing & 0 & 0 & 0 & 0\\
Total & 196 & 136 & 60 & 15\\
\bottomrule
\end{tabular}
\end{table}

The complete negative-energy inventory is \(5.4M_I\) (Table~\ref{tab:3}).
Excluding 28 primary crossings leaves \(4.4M_I\), including
\(3.5M_I\) of ice (79\%). There are no Saturn crossings. The tighter
run has the same row totals but exchanges one bound projectile-ice
element for one projectile-rock element, giving 79\% ice in the
no-crossing subset. Crossing flags affect reported inventories, not
the material's continued gravitational influence.

The primary's eccentricity increases slightly, remaining near 0.13
in both integrations, at semimajor axis near \(18R_S\). Its compact membership
explains the offset from the global first-impact remnant.
Figure~\ref{fig:titan-eccentricity} follows the continuing exchange
with debris.

The two integrations agree on the original selected population's
endpoint membership and on refractory departure, but differ by 530 m
s\(^{-1}\) in root-mean-square (RMS) unresolved velocities. Mechanical unbinding of the
wider fixed cohort retains 38 versus 41 members, and of the inner cohort
considered alone retains nine versus five. Both endpoint self-bound
subsets have positive Saturn energy. Energy conservation better than
\(3.1\times10^{-15}\) of the full system's initial mechanical energy
does not establish convergence of these internal trajectories (Appendix~\ref{appendix-c.-numerical-debris-controls-and-physical-fragment-assumptions}).

\subsubsection{Extended bound
trajectories}\label{extended-bound-trajectories}

Table~\ref{tab:4} compares the largest instantaneous semimajor axes in three
negative-energy populations. With \(\mathcal E_S\) the specific
Saturn-centered two-body energy, \(a=-GM_S/(2\mathcal E_S)\) uses
positions and velocities relative to recoiling Saturn. Both integrations
identify the same elements; tighter-run axes are
\(7.4\times10^3R_S\), \(310R_S\), and \(180R_S\).

\begin{table}[!htbp]
\centering
\caption{Maximum instantaneous Saturn-centered semimajor axes at five days in the
reference integration. Every entry has negative Saturn-centered two-body
energy. Compositions denote original material tags and body of origin.\label{tab:4}}
\small\setlength{\tabcolsep}{4pt}\renewcommand{\arraystretch}{1.15}
\begin{tabular}{@{}>{\raggedright\arraybackslash}p{(\linewidth - 6\tabcolsep)*\real{0.35000000}}>{\raggedright\arraybackslash}p{(\linewidth - 6\tabcolsep)*\real{0.15000000}}>{\raggedright\arraybackslash}p{(\linewidth - 6\tabcolsep)*\real{0.24000000}}>{\raggedright\arraybackslash}p{(\linewidth - 6\tabcolsep)*\real{0.26000000}}@{}}
\toprule
Population searched & Largest \(a/R_S\) & Eccentricity & Composition\\
\midrule
All negative-energy debris & \(7.4\times10^3\) & \(1-e\simeq2.1\times10^{-3}\) & Titan ice (100\%)\\
Original 97-element selection & 310 & 0.96 & Projectile rock (100\%)\\
Ice within that selection & 180 & 0.94 & Titan ice (100\%)\\
\bottomrule
\end{tabular}
\end{table}

The first entry lies at \(34R_S\), with pericenter \(16R_S\):
its large semimajor axis reflects a nearly parabolic orbit. Stability
requires solar perturbations, omitted here. The selected Titan-ice
sample evolves from \((a/R_S,e)=(290,0.96)\) at handoff to
\((180,0.94)\) after five days
(Figure~\ref{fig:iapetus-candidate-orbits}). These unresolved icy
trajectories supply starting states for assembly and damping.

\subsection{Initially bound companions: ice release and return
encounters}\label{initially-bound-companions-ice-release-and-return-encounters}

The three bound-satellite impacts reach 6,000 s. Each leaves all
projectile rock in a separate mechanically self-bound, Saturn-bound
survivor. Table~\ref{tab:5} reports only separated ice outside both large
remnants; the companion's much larger starting mantle is not
automatically available debris.

\begin{table}[!htbp]
\centering
\caption{Bound impacts from the common \(q=10R_S\), \(Q=30R_S\) companion orbit.
The contact speed \(u_{\rm imp}\) follows from the orbital approach. \(N_i\)
counts separated Saturn-bound ice elements; \(M_i\) is their mass. The
total energy \(E(t)\) includes kinetic, thermal, and gravitational
energies, and \(K_{\rm rel,contact}\) is the relative-motion kinetic energy
at contact. The final column gives \(\max|E(t)-E(0)|/K_{\rm rel,contact}\)
(Appendix~\ref{appendix-a.-numerical-controls-and-interpretation-of-the-pilot-results}).\label{tab:5}}
\small\setlength{\tabcolsep}{4pt}\renewcommand{\arraystretch}{1.15}
\begin{tabular}{@{}>{\raggedright\arraybackslash}p{(\linewidth - 12\tabcolsep)*\real{0.13000000}}>{\raggedright\arraybackslash}p{(\linewidth - 12\tabcolsep)*\real{0.11000000}}>{\raggedright\arraybackslash}p{(\linewidth - 12\tabcolsep)*\real{0.16000000}}>{\raggedright\arraybackslash}p{(\linewidth - 12\tabcolsep)*\real{0.09000000}}>{\raggedright\arraybackslash}p{(\linewidth - 12\tabcolsep)*\real{0.16000000}}>{\raggedright\arraybackslash}p{(\linewidth - 12\tabcolsep)*\real{0.14000000}}>{\raggedright\arraybackslash}p{(\linewidth - 12\tabcolsep)*\real{0.21000000}}@{}}
\toprule
\(m_p/M_T\) & Angle & \(u_{\rm imp}\) (km s\(^{-1}\)) & \(N_i\) & \(M_i/M_I\) & Titan \(e\) & Energy residual\\
\midrule
0.10 & \(60^\circ\) & 3.7 & 2 & 0.074 & 0.032 & 9.1\%\\
0.25 & \(60^\circ\) & 3.7 & 4 & 0.15 & 0.083 & 1.8\%\\
0.25 & \(45^\circ\) & 3.7 & 15 & 0.56 & 0.059 & 2.1\%\\
\bottomrule
\end{tabular}
\end{table}

All three separated-ice inventories are counts in these pilot
realizations, not converged efficiencies. Their global energy residuals
are comparable to or exceed the target surface-binding scale of that
small ice inventory; this comparison cannot localize the error
(Appendix~\ref{appendix-a.-numerical-controls-and-interpretation-of-the-pilot-results}).

The \(0.25M_T\), \(45^\circ\) case releases six target-ice and nine
companion-ice samples. Its outermost negative-energy ice sample has
\(a=85R_S\), \(e=0.89\), and \(q=9.7R_S\). No additional
mechanically bound spatial group is identified among the separated
samples, including groups below the usual reporting threshold. The
survivor still carries substantial ice and has acquired some of Titan's
mantle; it is not a bare core.

Table~\ref{tab:continuations} maps the bound-companion continuations.
Each handoff starts a separate gravitational calculation; the final two
rows extend the same saved 30-yr state.

\begin{table}[!htbp]
\centering
\caption{Guide to the orbital continuations of the \(0.25M_T\) bound
companion. Endpoint times are measured from the stated hydrodynamic
handoff. Contact calculations stop before a second hydrodynamic impact.
Numerical settings are in Appendix~\ref{appendix-a.-numerical-controls-and-interpretation-of-the-pilot-results}.\label{tab:continuations}}
\small\setlength{\tabcolsep}{4pt}\renewcommand{\arraystretch}{1.15}
\begin{tabular}{@{}>{\raggedright\arraybackslash}p{(\linewidth - 6\tabcolsep)*\real{0.18}}>{\raggedright\arraybackslash}p{(\linewidth - 6\tabcolsep)*\real{0.21}}>{\raggedright\arraybackslash}p{(\linewidth - 6\tabcolsep)*\real{0.29}}>{\raggedright\arraybackslash}p{(\linewidth - 6\tabcolsep)*\real{0.32}}@{}}
\toprule
Impact / handoff & Integration & Endpoint & Purpose\\
\midrule
\(60^\circ\); 3,000 and 6,000 s & Reference and tighter & Approach screen near 270 days & Establish a return collision course\\
\(45^\circ\); 5,000 s & Reference & Compact contact at 6.9 yr & Illustrative return and instantaneous merger estimate\\
\(45^\circ\); 5,000 s & Tighter & 10 yr, no contact & Check sensitivity of the return history\\
\(45^\circ\); 6,000 s & Reference & 30 yr, no contact & Follow Titan and the outer icy sample\\
\(45^\circ\); 6,000 s & Restart at 30 yr; 600-s step cap & Compact contact at 43 yr & Extend the reference trajectory\\
\(45^\circ\); 6,000 s & Three refined restarts at 30 yr & 50 yr, no contact & Test later encounter sensitivity (Table~\ref{tab:7})\\
\bottomrule
\end{tabular}
\end{table}

The larger \(60^\circ\) survivor reaches a collision course with Titan
near 270 days from both the 3,000-s and 6,000-s handoffs. The softened
pair-pericenter estimates, \(1.2\times10^3\) and \(2.5\times10^3\) km, lie inside both the
measured and compact-radius brackets. Estimated contact speeds are
3.6--3.7 km s\(^{-1}\), with contact angles about
\(23^\circ\)--\(26^\circ\) and \(40^\circ\)--\(48^\circ\) for the two
handoffs. Tighter orbital integrations retain this collision course. All
four separated ice samples remain Saturn bound without a recorded
central-body crossing up to the later-handoff approach. This establishes
a return geometry, before a second hydrodynamic impact.

\begin{figure}[!htbp]
\centering
\includegraphics[width=\linewidth,keepaspectratio,alt={Bound-companion orbital continuations. Top: minima of saved hourly center separations in one-day (left) and 30-day (right) bins; these curves do not resolve every closest approach. The 60-degree marker gives the independently calculated pair-pericenter estimate at its approach screen; the 45-degree star marks directly integrated compact contact. Dashed lines mark compact-radius sums. Lower left: separation-vector difference between the reference and tighter 5,000-s continuations at common saved times; the vertical line marks their first screened flyby. Lower right: separated ice without recorded central-body crossings at three distinct endpoints. These are different continuations of the two branches, not successive stages of one trajectory.}]{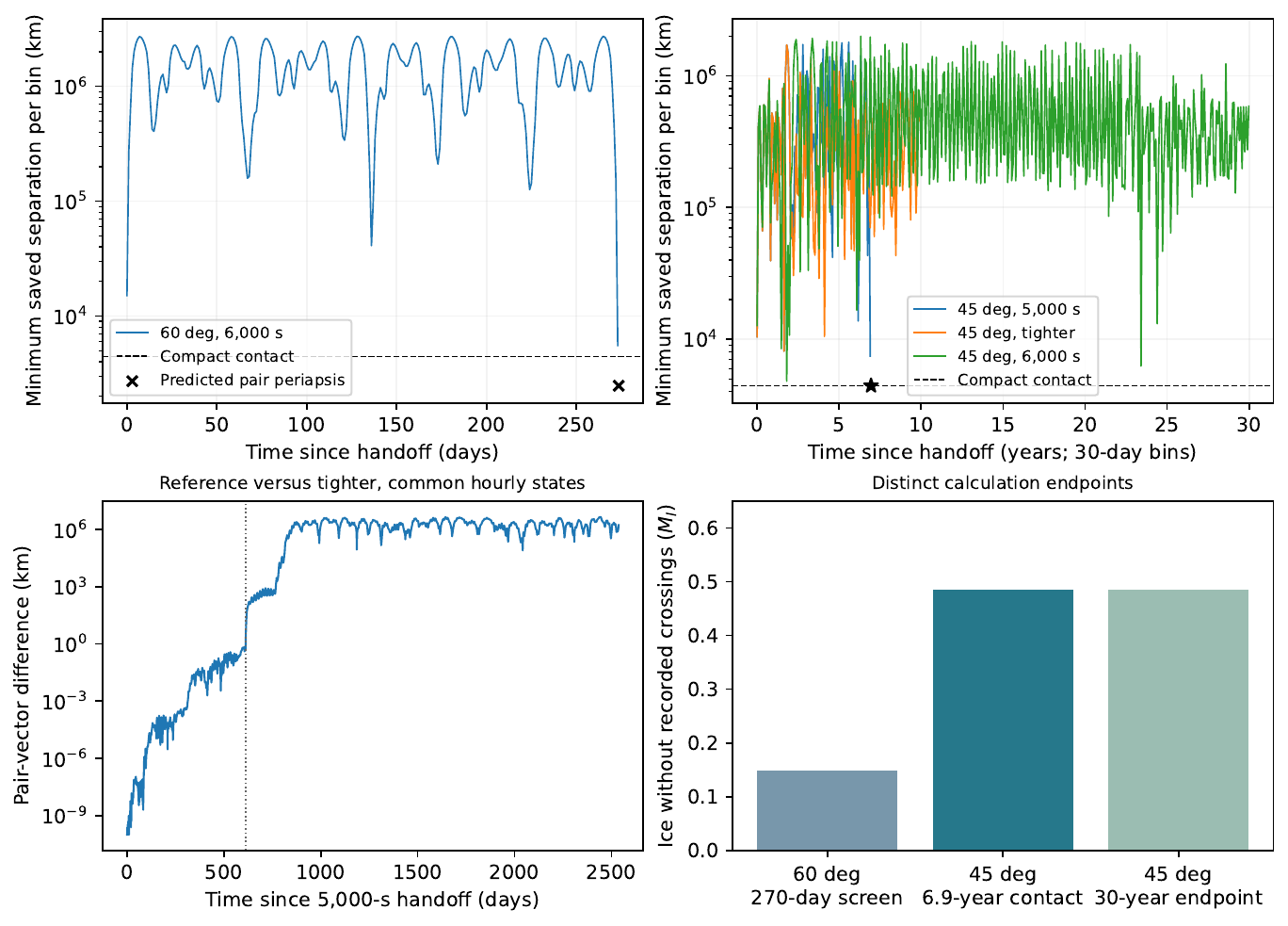}
\caption{Bound-companion orbital continuations. Top: minima of saved
hourly center separations in one-day (left) and 30-day (right) bins;
these curves do not resolve every closest approach. The \(60^\circ\)
marker gives the independently calculated pair-pericenter estimate at its
approach screen; the \(45^\circ\) star marks directly integrated compact
contact. Dashed lines mark compact-radius sums. Lower left:
separation-vector difference between the reference and tighter 5,000-s
continuations at common saved times; the vertical line marks their first
screened flyby. Lower right: separated ice without recorded central-body
crossings at three distinct endpoints. These are different continuations
of the two branches, not successive stages of one
trajectory.}\label{fig:bound-returns}
\end{figure}

For the \(45^\circ\) branch, one continuation from 5,000 s reaches
compact-body contact after 6.9 yr, at 3.1 km s\(^{-1}\) and
approximately \(50^\circ\) from head-on (Figure~\ref{fig:bound-returns}).
The actual separation reaches
the adopted radius sum of \(4.4\times10^3\) km. This continuation starts
with 16 separated icy samples at 5,000 s, one more Titan-ice sample than
the 15 counted at 6,000 s in Table~\ref{tab:5}. At contact all 16 retain negative
Saturn energy, totaling \(0.60M_I\); excluding three samples with
recorded primary crossings leaves 13, or \(0.48M_I\). The difference
from \(0.56M_I\) therefore reflects the earlier handoff, not growth of
the 6,000-s inventory. Its outermost sample has \(a=70R_S\),
\(e=0.86\), and \(q=9.6R_S\).

A complete-merger estimate from that saved contact state, with no new
ejecta or debris accretion, combines remnants of \(0.99M_T\) and
\(0.25M_T\). Denoting the merged remnant's mass, eccentricity, and
semimajor axis by \(M_2,e_2,a_2\), equation (26) gives
\(M_2=1.2M_T=1.7\times10^{23}\) kg, \(e_2=0.10\), and
\(a_2=19R_S\); Titan has \(e=0.11\) immediately before contact. A water-releasing second impact changes this mass and orbit through
the ejecta terms in equation (26).

The \(45^\circ\) continuation from 6,000 s has flybys near 570 and 670
days and no compact contact within 30 yr. It retains \(0.48M_I\)
of ice without central-body crossings, with outermost
\((a/R_S,e)=(60,0.84)\), while Titan reaches \(e_T=0.037\).
Section~\ref{continuation-beyond-thirty-years} follows the extension.
A tighter 5,000-s continuation has no contact within ten years.
Initially identical trajectories diverge after a shared flyby near day
610 with predicted pericenter \(7.1\times10^3\) km. The different
screening radii and hourly sampling limit interpretation of this
encounter sensitivity; Appendix~\ref{appendix-a.-numerical-controls-and-interpretation-of-the-pilot-results} gives the settings and timings.

The first collision supplies a partial reservoir. If Iapetus's eventual
ice fraction is \(f_{i,I}\) and the available ice mass is
\(M_{i,\mathrm{available}}\), the additional mass \(M_{i,\mathrm{additional}}\)
required satisfies

\[
M_{i,\mathrm{additional}}\geq
\max(0,f_{i,I}M_I-M_{i,\mathrm{available}}).
\tag{12b}
\]

For \(f_{i,I}=0.8\), the \(0.48M_I\) no-crossing inventory leaves at
least \(0.32M_I\) to supply before further losses. The returning
survivor could contribute; its initial mantle cannot also be counted
as free debris.

Completing this pathway requires a return impact that incorporates the
core while preserving or augmenting the ice. The calculations establish
return possibilities, but neither a converged \(45^\circ\) contact date
nor the second impact's outcome.

\FloatBarrier
\section{Assembly, scattering, and the angular-momentum
requirement}\label{assembly-scattering-and-the-angular-momentum-requirement}

\subsection{Growth competes with loss from the
reservoir}\label{growth-competes-with-loss-from-the-reservoir}

\citet{mosqueira2005} estimated growth and damping near Titan with debris
surface density \(\Sigma_d=200\) g cm\(^{-2}\). Their order-\(10^2\) yr
dynamical-friction scale could lengthen through stirring and gap opening.

Let \(\Omega_T=\sqrt{GM_S/a_T^3}\) be Titan's reference orbital angular
frequency. For a prospective precursor with radius \(r_b\), mean density
\(\rho_b\), and gravitational-focusing factor \(F_g\), the reference
accretion timescale \(t_{\rm acc}\) is

\[
t_{\rm acc}\sim\frac{\rho_b r_b}{\Sigma_d\Omega_TF_g}
\simeq\frac{3.4\times10^3}{F_g}\ \mathrm{yr}
\left(\frac{\rho_b}{1\ \mathrm{g\,cm^{-3}}}\right)
\left(\frac{r_b}{10^3\ \mathrm{km}}\right)
\left(\frac{\Sigma_d}{200\ \mathrm{g\,cm^{-2}}}\right)^{-1}.
\tag{13}
\]

The same proposal's scattering timescale \(t_{\rm sc}\) is

\[
t_{\rm sc}\sim\frac{0.1}{\Omega_T}
\left(\frac{M_S}{M_T}\right)^2
\simeq1.2\times10^4\ \mathrm{yr}.
\tag{14}
\]

Their \(10^3\)--\(10^4\) yr collisional-removal estimate for Titan-crossing
bodies overlaps these scales. Applying them to the pilot debris requires
its evolving density and velocity distribution.

\subsection{A distant apocenter is not
emplacement}\label{a-distant-apoapsis-is-not-emplacement}

\begin{figure}[!htbp]
\centering
\includegraphics[width=\linewidth,keepaspectratio,alt={Mass-bearing orbital distribution of all 145 negative-energy unresolved elements at five days in the reference unbound-impact continuation. Histograms retain original body and material identity and include the 28 elements with recorded primary crossings; their separate inventory is in Table 3. The middle panel plots p = a(1 - e squared), the circularization radius at fixed specific angular-momentum magnitude h. The dotted lines mark 20 Saturn radii and the dashed line marks 60 Saturn radii. Inclination is measured from the initial prograde impact plane. All bins contain numerical mass weights, not counts of resolved physical bodies.}]{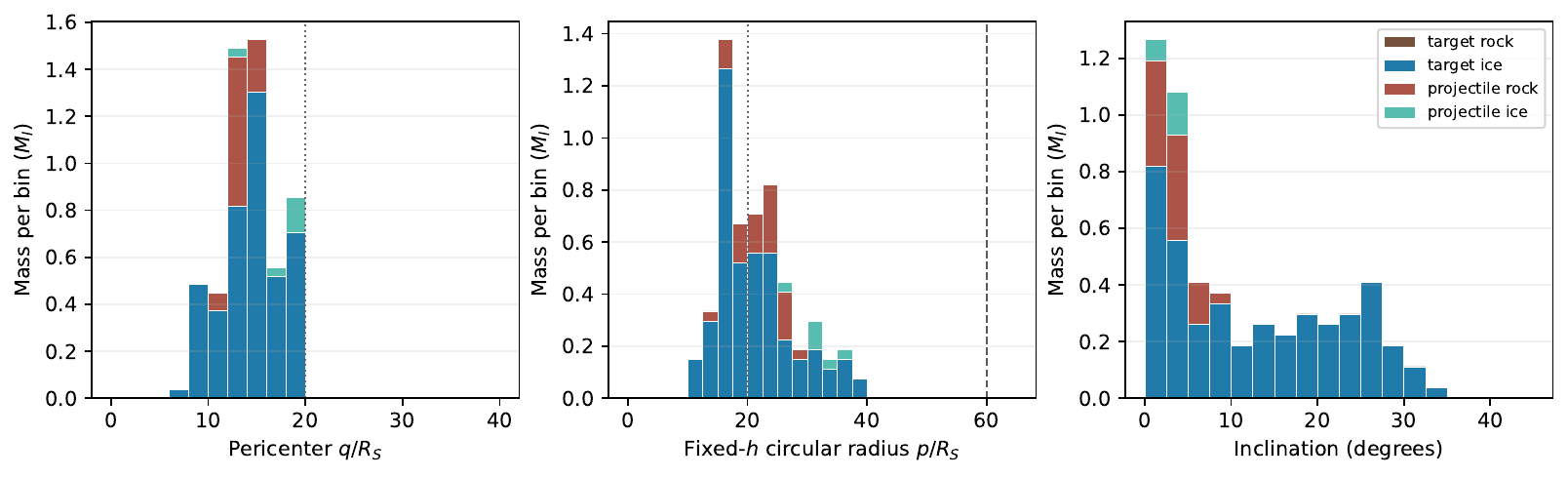}
\caption{Mass-bearing orbital distribution of all 145 negative-energy
unresolved elements at five days in the reference unbound-impact
continuation. Histograms retain original body and material identity and
include the 28 elements with recorded primary crossings; their separate
inventory is in Table~\ref{tab:3}. The middle panel plots \(p=a(1-e^2)\), the
circularization radius at fixed specific angular-momentum magnitude \(h\). The dotted
lines mark \(20R_S\) and the dashed line marks \(60R_S\). Inclination is
measured from the initial prograde impact plane. All bins contain
numerical mass weights, not counts of resolved physical
bodies.}\label{fig:debris-orbits}
\end{figure}

Figure~\ref{fig:debris-orbits} shows the complete negative-energy unresolved inventory after
the unbound pilot's five-day continuation. In the reference run, all 145
elements are prograde, with inclinations
\(0.058^\circ\)--\(35^\circ\), pericenters \(8.0\)--\(20R_S\),
and semilatus recta \(p=a(1-e^2)=11\)--\(40R_S\). The tighter run has the same count
and prograde orientation, with corresponding ranges
\(0.058^\circ\)--\(31^\circ\), \(6.9\)--\(20R_S\), and
\(9.6\)--\(40R_S\). This is an extended, eccentric population
crossing Titan's orbital region, not an already detached circular outer
disk.

For a Saturn-bound orbit with semimajor axis \(a\), eccentricity \(e\),
pericenter \(q\), apocenter \(Q\), and specific angular-momentum
magnitude \(h\),

\[
q=a(1-e),\qquad Q=a(1+e),\qquad
h^2=GM_Sa(1-e^2)=GM_S\frac{2qQ}{q+Q}.
\tag{15}
\]

An orbit scattered near radius \(r_{\rm enc}\) must still pass through
that encounter region. For \(a=60R_S\) and an encounter near \(20R_S\),
this requires \(e\gtrsim2/3\). Reducing its eccentricity at fixed
specific angular momentum would leave a circular orbit of radius
\(a_{\rm circ}\):

\[
a_{\rm circ}=\frac{h^2}{GM_S}=a(1-e^2)=q(1+e).
\tag{16}
\]

More generally, any bound orbit passing through \(r_{\rm enc}\)
satisfies \(h^2<2GM_Sr_{\rm enc}\), because its speed is below local
escape speed and its transverse speed cannot exceed its total speed.
Consequently,

\[
a_{\rm circ}<2r_{\rm enc}.
\tag{17}
\]

\begin{figure}[!htbp]
\centering
\includegraphics[width=0.84\linewidth,keepaspectratio,alt={Angular-momentum constraint for a bound post-encounter orbit. Below the solid boundary, fixed-h circularization can end at radii smaller than twice the encounter radius; radii above it require additional angular momentum or a later encounter farther out. The horizontal line marks the 60 Saturn radii target. An encounter at 20 Saturn radii gives an upper limit of 40 Saturn radii, irrespective of how far the initial apocenter extends. This figure is calculated from equation (17), not from the impact particles.}]{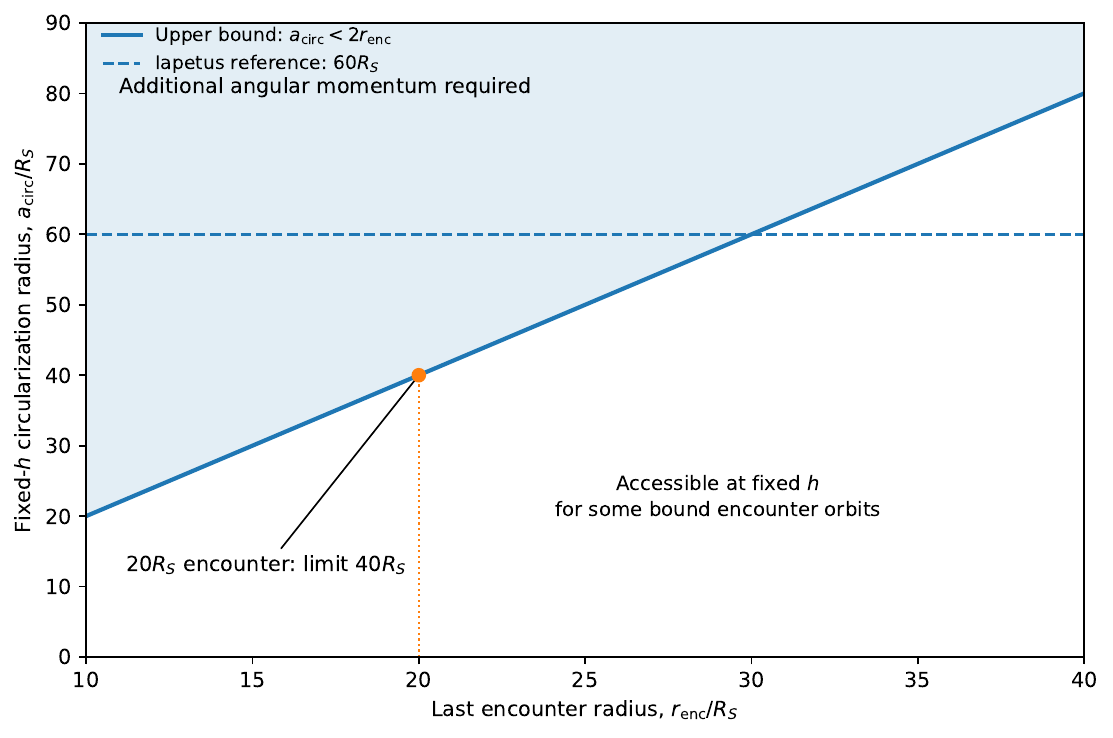}
\caption{Angular-momentum constraint for a bound post-encounter orbit.
Below the solid boundary, fixed-\(h\) circularization can end at radii
smaller than \(2r_{\rm enc}\); radii above it require additional angular
momentum or a later encounter farther out. The horizontal line marks the
\(60R_S\) target. An encounter at \(20R_S\) gives an upper limit of
\(40R_S\), irrespective of how far the initial apocenter extends. This
figure is calculated from equation (17), not from the impact
particles.}\label{fig:angular-momentum}
\end{figure}

An encounter at \(20R_S\) followed solely by fixed-\(h\) energy loss
therefore circularizes inside \(40R_S\). A circular orbit at \(60R_S\)
requires at least a 22\% increase in specific angular momentum
relative to even the limiting bound encounter orbit. Figure~\ref{fig:angular-momentum} shows this
general geometric constraint. The encounter radius is the actual last
encounter location; an eccentric Titan can encounter the precursor
farther out than its initial semimajor axis.

For comparison, \((q,Q)=(20,60)R_S\) gives \(a=40R_S\), \(e=0.5\),
and \(a_{\rm circ}=30R_S\); \((a,e)=(60R_S,0.7)\) gives
\(q=18R_S\), \(Q=100R_S\), and \(a_{\rm circ}=31R_S\).
A distant apocenter alone therefore does not specify emplacement.

For any of these starting states, the specific-angular-momentum increase
\(\Delta h\) needed for a circular orbit at \(60R_S\) is

\[
\frac{\Delta h}{h}=\sqrt{\frac{60R_S}{a_{\rm circ}}}-1.
\tag{18}
\]

Table~\ref{tab:6} gives the values for the reference trajectories.

\begin{table}[!htbp]
\centering
\caption{Orbital accounting for the reference trajectory elements. Values are
analytical evaluations rounded for display, not new integrations.
Calculations retain the reference elements at their archived precision. The last column gives the
required increase in the magnitude of specific angular momentum; the
prograde orientation and inclination must also be followed.\label{tab:6}}
\small\setlength{\tabcolsep}{4pt}\renewcommand{\arraystretch}{1.15}
\begin{tabular}{@{}>{\raggedright\arraybackslash}p{(\linewidth - 10\tabcolsep)*\real{0.29000000}}>{\raggedright\arraybackslash}p{(\linewidth - 10\tabcolsep)*\real{0.13000000}}>{\raggedright\arraybackslash}p{(\linewidth - 10\tabcolsep)*\real{0.14500000}}>{\raggedright\arraybackslash}p{(\linewidth - 10\tabcolsep)*\real{0.13000000}}>{\raggedright\arraybackslash}p{(\linewidth - 10\tabcolsep)*\real{0.16000000}}>{\raggedright\arraybackslash}p{(\linewidth - 10\tabcolsep)*\real{0.14500000}}@{}}
\toprule
Starting state & \(a/R_S\) & \(e\) & \(q/R_S\) & \(a_{\rm circ}/R_S\) & \(\Delta h/h\)\\
\midrule
Historical illustration & 60 & 0.70 & 18 & 31 & 40\%\\
Unbound impact, selected ice & 180 & 0.94 & 11 & 21 & 67\%\\
Bound \(45^\circ\), 6,000 s & 85 & 0.89 & 9.7 & 18 & 81\%\\
Bound \(45^\circ\), return & 70 & 0.86 & 9.6 & 18 & 83\%\\
Bound \(45^\circ\), 30 yr & 60 & 0.84 & 9.4 & 17 & 86\%\\
\bottomrule
\end{tabular}
\end{table}

The measured trajectories require substantially more fractional angular
momentum than the historical illustration, yet the system's scale
remains adequate. For one Iapetus mass, the total orbital
angular-momentum increase \(\Delta L_I=M_I\Delta h\) satisfies

\[
\frac{\Delta L_I}{M_T\sqrt{GM_S(20R_S)}}
=\frac{M_I}{M_T}\left[\sqrt{3}-\sqrt{\frac{a_{\rm circ}}{20R_S}}\right].
\tag{19}
\]

The selected \(180R_S\) ice sample and return-contact example require
0.94\% and 1.1\% of Titan's reference orbital angular momentum; the
historical illustration requires 0.66\%. These are magnitude comparisons
for aligned transfer. The actual exchange must also establish the
orbital plane.

\FloatBarrier
\section{Damping by debris and residual gas}\label{can-the-debris-circularize-iapetus-or-is-gas-required}

\subsection{Debris as both building material and a dynamical
agent}\label{debris-as-both-building-material-and-a-dynamical-agent}

Gravitational encounters exchange energy and angular momentum; accretion
and inelastic collisions additionally dissipate relative motion.
Planetesimal disks can recircularize scattered low-mass planets
\citep{raymond2010}. Applying this process to correlated collision debris
requires following the reservoir's changing mass and orbits.

The angular-momentum budget sets a useful scale independently of a
detailed friction formula. Let \(h_i\) and \(h_f\) be the precursor's
initial and final specific angular-momentum magnitudes. If it requires
\(\Delta L_I=M_I(h_f-h_i)\) and a donor mass \(M_d\) loses a mean
specific angular momentum \(\Delta h_d\), then

\[
M_d\Delta h_d\geq\Delta L_I.
\tag{20}
\]

For the historical \((a,e)=(60R_S,0.7)\) precursor, perfect transfer from
material moving between circular orbits at \(60R_S\) and \(20R_S\)
requires \(0.68M_I\) of donor mass. This illustrates the available
angular-momentum scale, without establishing transfer efficiency.

Impact debris reaches large radii on eccentric orbits and has less
angular momentum than circular material there. Treating it as a cold
outer disk would add an uncalculated reservoir. Its redistribution,
reaccretion, escape, and assembly must instead be evolved together,
with equation (3) preventing double counting.

The reciprocal-gravity runs include debris exchange; the original
one-way tracers do not. Neither treatment includes the collisions and
growth needed to test debris-assisted assembly and circularization.

\subsection{Why the force must act relative to the rotating
medium}\label{why-the-force-must-act-relative-to-the-rotating-medium}

For Saturn-centered position \(\boldsymbol r\) and velocity
\(\boldsymbol v\), let \(\mathcal E=|\boldsymbol v|^2/2-GM_S/r\) be
the specific two-body energy and \(\boldsymbol h=\boldsymbol r\times\boldsymbol v\)
the specific angular momentum, with \(r=|\boldsymbol r|\) and
\(h=|\boldsymbol h|\). Any additional acceleration \(\boldsymbol a_d\)
changes them according to (dots denote time derivatives)

\[
\dot{\mathcal E}=\boldsymbol v\cdot\boldsymbol a_d,\qquad
\dot{\boldsymbol h}=\boldsymbol r\times\boldsymbol a_d.
\tag{21}
\]

Writing \(e^2=1+2\mathcal Eh^2/(GM_S)^2\) gives

\[
\frac{d e^2}{dt}=
\frac{2h^2}{(GM_S)^2}\dot{\mathcal E}
+\frac{4\mathcal E}{(GM_S)^2}\boldsymbol h\cdot\dot{\boldsymbol h}.
\tag{22}
\]

Together with \(a=-GM_S/(2\mathcal E)\), these relations couple
eccentricity damping to semimajor-axis and angular-momentum evolution.
Successful emplacement requires raising pericenter away from Titan
without excessive orbital contraction.

The sign can be seen in a coplanar circular comparison medium. With
semilatus rectum \(p=a(1-e^2)\), true anomaly \(f\), azimuthal speed
\(v_\phi\), and local circular speed \(v_K(r)=\sqrt{GM_S/r}\),

\[
\frac{v_\phi}{v_K(r)}=\sqrt{\frac{p}{r}}
=\sqrt{1+e\cos f}.
\tag{23}
\]

Near pericenter the body moves faster azimuthally than circular material
and tends to lose angular momentum; near apocenter a faster medium can
add it. Both dissipate relative motion, with different signs of work in
the Saturn frame. The net exchange depends on the medium's extent,
density, and velocity distribution.

\subsection{Residual gas: a plausible additional
reservoir}\label{residual-gas-a-plausible-additional-reservoir}

Residual circumplanetary gas could damp a precursor directly or through
smaller debris. Impact vapor is a separate reservoir whose mass, extent,
and angular momentum follow from the collision.

For a body of physical radius \(s_b\) and density \(\rho_b\), let
\(C_D\) be the dimensionless drag coefficient, \(\rho_g\) the gas density,
and \(\boldsymbol v_g\) its Saturn-centered velocity. The relative
velocity \(\boldsymbol u\) and aerodynamic acceleration
\(\boldsymbol a_{\rm aero}\) are \citep{adachi1976}

\[
\boldsymbol a_{\rm aero}=-\frac{3C_D\rho_g}{8\rho_b s_b}
|\boldsymbol u|\boldsymbol u,\qquad
\boldsymbol u=\boldsymbol v-\boldsymbol v_g.
\tag{24}
\]

Gas density and rotational velocity must be consistent with pressure
support. Equation (24) requires physical fragment sizes; SPH smoothing
lengths do not determine aerodynamic area-to-mass ratios.

A local calculation illustrates the magnitude without assigning a disk
solution. Denote gas temperature by \(T_g\), isothermal sound speed by
\(c_s\), vertical density scale height by \(H\), gas surface density by
\(\Sigma_g\), and local orbital angular frequency by
\(\Omega=\sqrt{GM_S/r^3}\). At \(60R_S\), let a vertically isothermal
molecular gas have
\(T_g=90\) K and mean molecular weight 2.3. Then \(c_s\simeq560\) m
s\(^{-1}\) and \(H=c_s/\Omega\simeq6.2\times10^8\) m. A Gaussian column
\(\Sigma_g=100\) g cm\(^{-2}\) gives midplane density
\(6.4\times10^{-7}\) kg m\(^{-3}\). For an Iapetus-sized body with
\(s_b=730\) km, \(\rho_b=1.1\) g cm\(^{-3}\), and relative speed 1 km
s\(^{-1}\), let \(u=|\boldsymbol u|\). Equation (24) gives the local
speed-damping time \(t_u\):

\[
t_u\equiv\frac{u}{|\boldsymbol a_{\rm aero}|}
\simeq1.1\times10^5\ \mathrm{yr}
\left(\frac{C_D}{1}\right)^{-1}
\left(\frac{\Sigma_g}{100\ \mathrm{g\,cm^{-2}}}\right)^{-1}
\left(\frac{u}{1\ \mathrm{km\,s^{-1}}}\right)^{-1},
\tag{25}
\]

at the stated body and gas properties. A 10 g cm\(^{-2}\) column gives
ten times this scale. These columns match the residual-gas benchmarks
of \citet{mosqueira2005}; equation (25) is a local speed-damping estimate,
not an orbit-integrated eccentricity timescale. Relative velocity and
sampled gas density vary around the orbit and with inclination.

For moving-body mass \(M_b\) and dimensionless wake factor \(\mathcal I\),
set by Mach number and geometry, the homogeneous-medium force scale is
\(4\pi(GM_b)^2\rho_g\mathcal I/u^2\) \citep{ostriker1999}.
Application to an eccentric satellite requires the finite, shearing
disk geometry.

Gas could maintain damping after solid depletion, but concentrated inner
braking can remove angular momentum and draw the precursor inward.
Outer retention depends on the net torque.

\subsection{Return impacts and subsequent orbital evolution}\label{refractory-escape-or-return-and-titans-final-eccentricity}

A return impact changes Titan's mass and momentum.
Let \(M_1\) and \(m_c\) be the approaching primary and companion masses,
with Saturn-centered velocities \(\boldsymbol v_1^-\) and
\(\boldsymbol v_c^-\) immediately before contact. Let \(m_{\rm ej}\) and
\(\boldsymbol p_{\rm ej}\) be the mass and momentum of all material left
outside the new dominant remnant, whose mass and velocity are \(M_2\) and
\(\boldsymbol v_2\). Neglecting the external impulse during a short collision,

\[
\begin{aligned}
M_2\boldsymbol v_2&=M_1\boldsymbol v_1^-+m_c\boldsymbol v_c^--\boldsymbol p_{\rm ej},\\
M_2&=M_1+m_c-m_{\rm ej}.
\end{aligned}
\tag{26}
\]

The remnant position follows from the same mass distribution, and
equation (9) gives its orbit. Spin and ejecta angular momentum must also
be accounted for. Section~\ref{initially-bound-companions-ice-release-and-return-encounters} reports the zero-ejecta merger estimate.

\begin{figure}[!htbp]
\centering
\includegraphics[width=\linewidth,alt={Titan eccentricity histories for the unbound-impact and bound-companion scenarios, including debris gravity and remnant recoil.}]{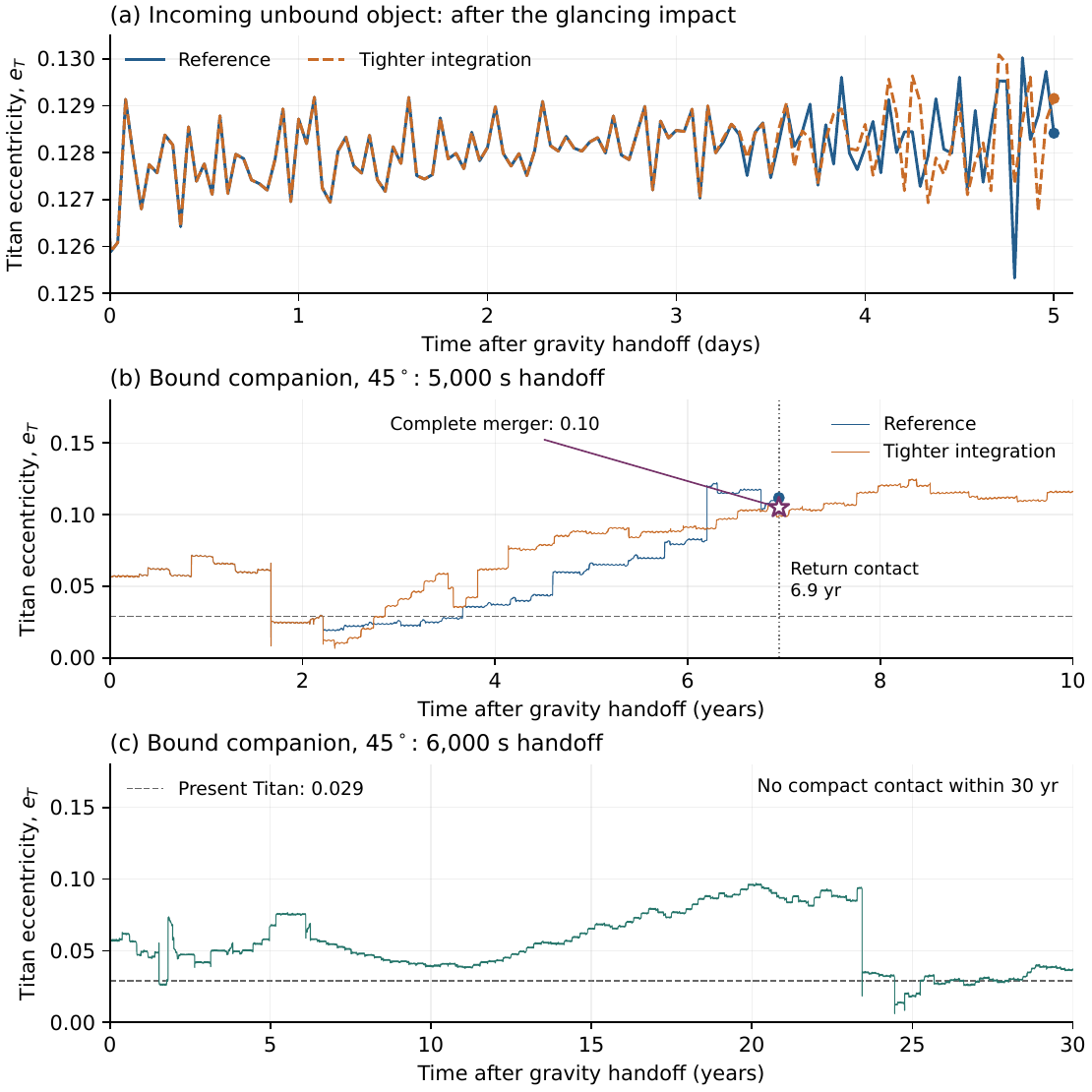}
\caption{Titan's Saturn-centered osculating eccentricity in both collision
scenarios. All unresolved debris gravitates reciprocally, and Saturn and
Titan recoil; gas damping is absent. (a) Five days after the 3,000-s
unbound-impact handoff, with an expanded vertical scale. (b) The
\(45^\circ\), \(0.25M_T\) bound companion, followed from 5,000 s in the
reference and tighter integrations. The filled circle marks Titan at
compact return contact after 6.9 yr; the open star is the instantaneous
complete-merger estimate \(e_T=0.10\), not an evolved second impact.
(c) The same bound collision followed from 6,000 s for 30 yr, without
compact contact. Each time axis starts at its stated handoff. Dashed
horizontal lines mark present Titan's \(e_T\simeq0.029\). Curves use
all saved samples without time averaging. Figure~\ref{fig:continuation-steps}
extends panel (c) beyond 30 yr.\label{fig:titan-eccentricity}}
\end{figure}

\begin{figure}[!htbp]
\centering
\includegraphics[width=0.92\linewidth,alt={Semimajor axis and eccentricity histories of fixed Iapetus-candidate ice samples in the unbound-impact and bound-companion scenarios.}]{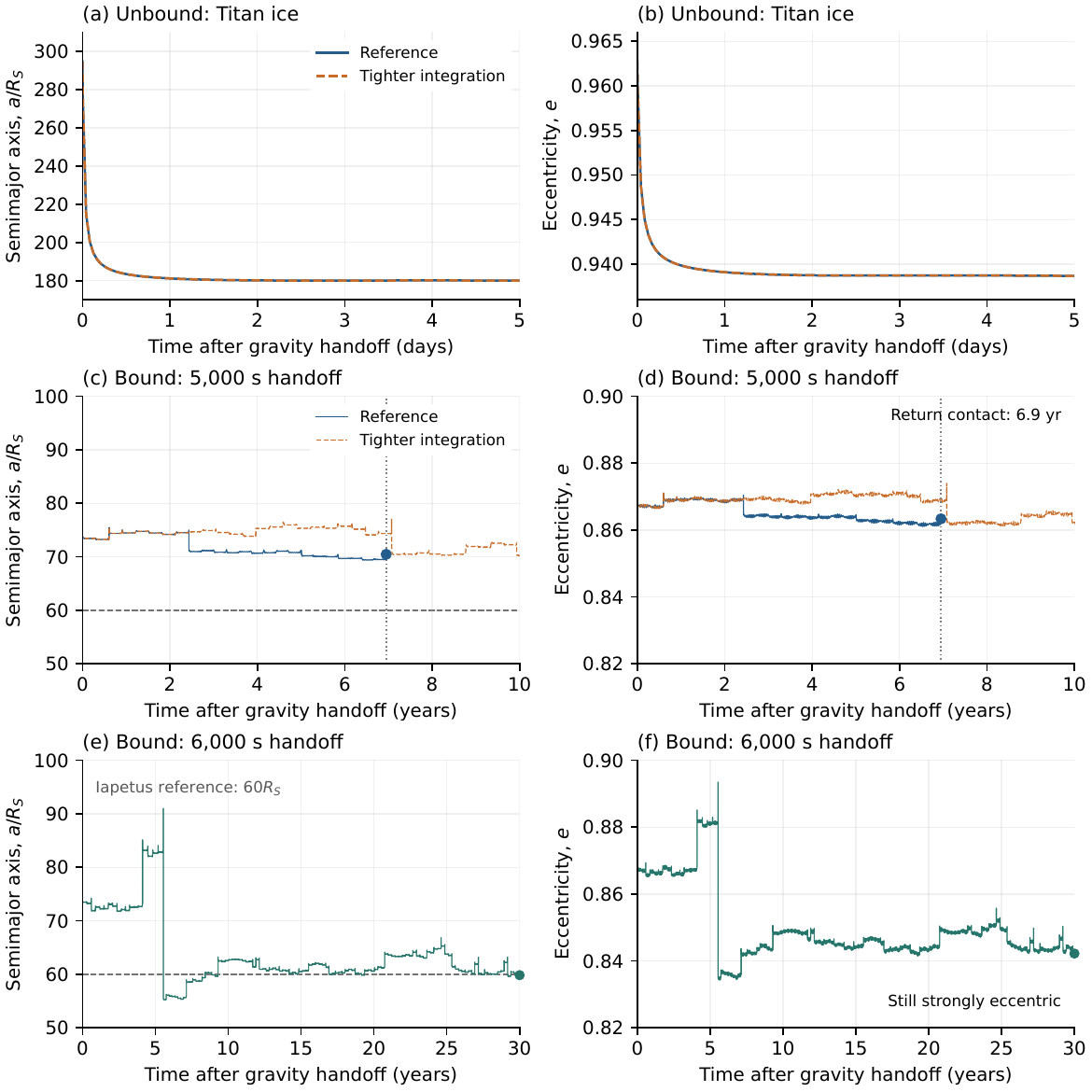}
\caption{Orbital evolution of the candidate icy material in the same
continuations as Figure~\ref{fig:titan-eccentricity}. Left: Saturn-centered
osculating semimajor axis; right: eccentricity. Each curve follows the
same ice sample at every saved time. The top row follows Titan ice in
the unbound-impact scenario; the middle and bottom rows follow the same
companion-ice sample from the \(45^\circ\) bound collision, starting
at 5,000 and 6,000 s, respectively. Reference and tighter unbound curves
nearly coincide. Filled circles and vertical dotted lines in the middle
row mark the reference return contact at 6.9 yr; that trajectory ends
there. Horizontal dashed lines mark the \(60R_S\) Iapetus reference.
Debris gravity and remnant recoil are included; gas forces and icy
assembly are not evolved. The bottom-row continuation and its numerical
comparisons appear in Figure~\ref{fig:continuation-steps}.\label{fig:iapetus-candidate-orbits}}
\end{figure}

Figures~\ref{fig:titan-eccentricity} and~\ref{fig:iapetus-candidate-orbits}
collect the orbital histories for the results in Sections~\ref{refractory-departure-the-complete-inventory-and-titans-excitation}--\ref{initially-bound-companions-ice-release-and-return-encounters};
Table~\ref{tab:continuations} identifies the bound continuations.
Debris scattering and companion encounters continue to change Titan's
eccentricity after impact, as anticipated by \citet{mosqueira2005}.
The icy samples approach Iapetus's orbital scale while retaining high
eccentricity.

\subsubsection{Continuation beyond thirty years}\label{continuation-beyond-thirty-years}

These long continuations adopt the pressure-free handoff and compact
remnant radii of Section~\ref{a-glancing-collision-with-a-bound-satellite}; no cooling history is calculated.
Resuming the 6,000-s handoff from its saved 30-yr state gives
Titan--companion contact at 43 yr with the original 600-s step cap.
The contact speed is 2.7 km s\(^{-1}\), the angle is
\(48^\circ\) from head-on, and Titan has instantaneous
\(e_T=0.18\). No second impact or merger is evolved. The tracked icy
sample remains bound at \(a=56R_S\), \(e=0.83\).

\begin{figure}[!htbp]
\centering
\includegraphics[width=\linewidth,alt={Four continuations from the same thirty-year state: icy sample semimajor axis and eccentricity, and Titan eccentricity. The original integration reaches contact; three smaller-step comparisons reach fifty years without contact.}]{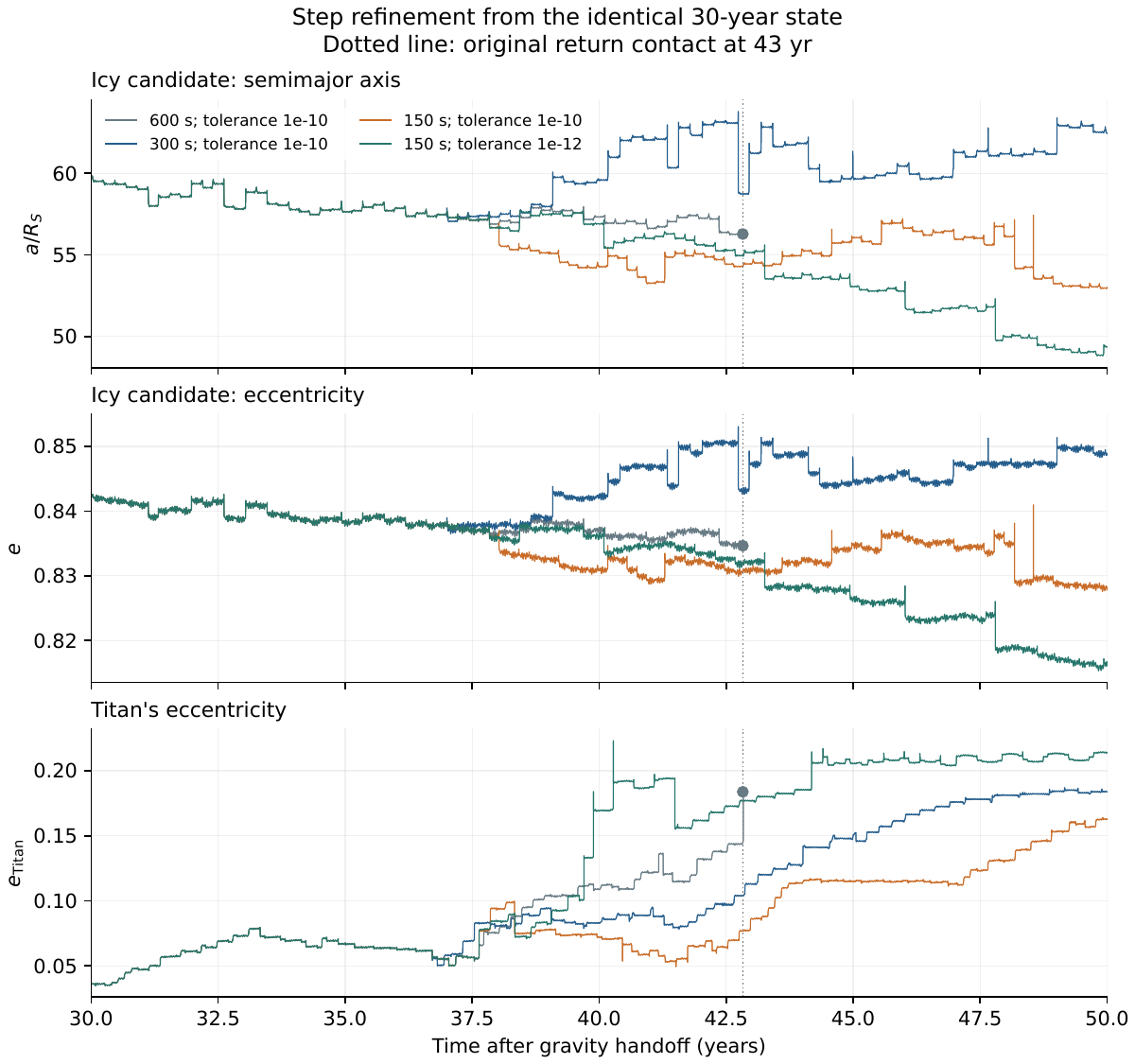}
\caption{Evolution beyond the bottom panels of
Figures~\ref{fig:titan-eccentricity} and~\ref{fig:iapetus-candidate-orbits}.
The four curves start from the identical saved 30-yr state, with the
step caps and IAS15 tolerances in Table~\ref{tab:7}. The dotted line and
endpoint markers identify Titan--companion contact in the original
600-s continuation; that curve stops before the second collision.
All curves include reciprocal debris gravity and remnant recoil, with
no gas damping or icy assembly. The candidate remains bound and highly
eccentric throughout.\label{fig:continuation-steps}}
\end{figure}

Three restarts from the identical 30-yr state, with smaller step caps
and otherwise the same forces and contact radii, reach 50 yr without
compact-body contact (Table~\ref{tab:7}; Figure~\ref{fig:continuation-steps}).
These comparisons leave the earlier history fixed; they do not establish
a converged return date or permanent avoidance of collision. All retain
the same companion-ice sample, of mass \(0.037M_I\), without a recorded
Saturn or Titan crossing. It represents unresolved ice, not an assembled
Iapetus. Its endpoint pericenter remains \(9.1\)--\(9.4R_S\): the
lower eccentricities mostly accompany orbital contraction, rather than
separation from Titan's orbital region. Their fixed-\(h\)
circularization radii are \(16\)--\(17R_S\); reaching \(60R_S\)
requires 86--91\% additional specific angular momentum. Titan's low
eccentricity at 30 yr is also temporary; subsequent scattering raises it to
\(0.16\)--\(0.21\) at the tabulated endpoints. Across the full unresolved
population, these endpoints retain \(0.45\)--\(0.48M_I\) of ice with
negative Saturn-centered energy and no recorded central-body crossing:
still a partial reservoir, rather than mass assembled on the tracked
outer orbit.

\begin{table}[!htbp]
\centering
\caption{Continuations from the same 30-yr state. Times are measured from
the 6,000-s gravity handoff. The first row ends at compact Titan--companion
contact; the others end at 50 yr without contact. Orbital elements are
instantaneous endpoint values for the same icy sample and for Titan;
\(e_T\) in the contact row precedes any merger. Times and orbital elements
are rounded independently from full precision.\label{tab:7}}
\small\setlength{\tabcolsep}{5pt}
\begin{tabular}{@{}rrrrrrr@{}}
\toprule
Step cap (s) & IAS15 tolerance & Time (yr) & \(a/R_S\) & \(e\) & \(q/R_S\) & \(e_T\)\\
\midrule
600 & \(10^{-10}\) & 43 & 56 & 0.83 & 9.3 & 0.18\\
300 & \(10^{-10}\) & 50 & 62 & 0.85 & 9.4 & 0.18\\
150 & \(10^{-10}\) & 50 & 53 & 0.83 & 9.1 & 0.16\\
150 & \(10^{-12}\) & 50 & 49 & 0.82 & 9.1 & 0.21\\
\bottomrule
\end{tabular}
\end{table}

\FloatBarrier
\subsubsection{Close encounters and circularization}\label{close-encounters-and-circularization}

An impulse estimate illustrates the competition between strong scattering
and collision. Place the 30-yr icy orbit, \(a=60R_S\) and
\(e=0.84\), at a coplanar prograde crossing of a circular Titan at
\(20R_S\). The asymptotic relative encounter speed is
\(u=5.0\) km s\(^{-1}\), approximated by the unperturbed crossing
velocities. For Titan's
modeled remnant mass \(M_1=1.3\times10^{23}\) kg and condensed radius
\(R_1\simeq2.8\times10^3\) km, consider a physical Iapetus-mass body with radius
730 km. This is an assumed encounter geometry, not a measured flyby or
a mass assigned to the numerical ice sample. If \(d\) is the closest
center separation, \(b_\infty\) the asymptotic impact parameter, and
\(\alpha=G(M_1+M_I)/u^2\), the icy body's velocity change
\(\Delta\boldsymbol v\) from Newtonian two-body deflection, including
Titan's recoil, is

\[
|\Delta\boldsymbol v|=\frac{2GM_1}{u(d+\alpha)},\qquad
b_\infty^2=d^2+2\alpha d.
\tag{27}
\]

Recomputing the Saturn-centered orbit over encounter orientations shows
that a decrease in eccentricity of 0.05 requires
\(d\lesssim2.0R_1\), compared with physical contact at \(1.3R_1\).
Including focusing, the available noncollisional area is at most
about 1.4 times the collision area for a locally uniform incoming flux; only
some orientations give the required decrease. This is a conditional
cross-section comparison, not a circularization probability. At
\(d=2R_1\), the favorable kick lowers eccentricity by 0.050 but also
contracts the orbit to \(a=37R_S\) and lowers pericenter from
\(9.4R_S\) to \(7.6R_S\). The reduced eccentricity therefore does not
detach the orbit from Titan's region. Weaker cumulative encounters
remain possible; the geometry and relative speed matter.

For circular Titan at \(20R_S\) and fixed \(a=60R_S\), \(e<2/3\)
eliminates geometric orbit crossing, not every finite-distance
encounter. Suppression of strong encounters additionally requires
clearance from Titan's encounter region and its evolving radial
excursion. The trajectories in Table~\ref{tab:7} remain crossing. As a
retained outer orbit detaches, strong scattering becomes less effective
at completing circularization. Within the proposed pathway, an early,
substantial collisional debris disk, residual gas, or both would supply
continued damping and angular-momentum exchange.

\FloatBarrier
\section{Discussion}\label{discussion}

We test whether collisions can supply an icy debris reservoir 
and follow the
initial stages of its orbital evolution. Iapetus's origin also requires
the debris to assemble and settle onto a
stable distant orbit. A gas-assisted completion of
the model would need to include the effects of
residual gas. Conceivably, a 
residual gas disk could be used to accomplish five tasks:

\textbf{Circularize Titan.} Damp the primary's eccentricity from its 
post-impact value.

\textbf{Clear the outer debris.} Remove the diffuse
icy population through aerodynamic gas drag \citep{weidenschilling1977} or
reaccretion,
while preserving material that assembles into Iapetus.

\textbf{Supply ice inward.} Deliver and retain some debris in the inner
disk, potentially explaining the ice-rich inner satellites and rings.

\textbf{Circularize and retain Iapetus.} Raise Iapetus's 
pericenter and retain
the precursor near \(60R_S\), stranding it in the outer disk 
through the clearing of other ejecta material by gas drag.

\textbf{Capture Hyperion into resonance with Titan.} 
One can view Hyperion
as a collisional remnant \citep{mosqueira2014} and gas drag
could subsequently capture it into resonance.

We do not attempt to model all of 
these processes here, nor have we demonstrated
that all these observations should be explained by a unified scenario. 
In addition, there
is also the issue that the collisional eccentricity excitation would need
to fit with the subsequent tidal eccentricity evolution, which is
difficult to reconcile with an early impact in the presence of residual
gas, if tidal damping remained efficient after gas dispersal
\citep{petricca2025}.

\FloatBarrier
\section{Conclusions}\label{conclusions}

The requirements for a collisional origin of Iapetus are sufficient
water-rich ejecta and a route that carries enough of it to a stable distant
orbit. The unbound impacts release several Iapetus masses of predominantly
Titan-derived, ice-tagged material into Saturn-bound trajectories.
In the oblique case, a \(0.10M_T\) impactor strikes at a speed of 10 km s\(^{-1}\).
About \(3.5M_I\) of the initially selected material
remains Saturn-bound after five days and is 83\% ice-tagged. An icy sample has an
orbit extending beyond Iapetus's distance, while a refractory projectile
concentration moves outward on a hyperbolic trajectory. The primary in the
oblique unbound case has an eccentricity near 0.13. 

In the bound
branch, a \(0.25M_T\) companion striking at \(45^\circ\) and about
3.7 km s\(^{-1}\) separates \(0.60M_I\) of ice and leaves a survivor.
One reference continuation reaches return contact after
6.9 yr while retaining \(0.48M_I\) of ice without recorded primary
crossings. A no-ejecta perfect merger gives
Titan an eccentricity near 0.10 following the merger.
The return histories are numerically sensitive, and continuations
without a merger can re-excite Titan's eccentricity.

After impact, debris trajectories remain highly eccentric and cross Titan's orbit. Therefore, forming
Iapetus requires survival and assembly of enough icy material, as well as
a way to raise the pericenter of the ensuing icy satellite.
Debris and residual gas could provide the exchanges while also damping
Titan's eccentricity, but their coupled evolution has not been treated.

\begin{acknowledgments}
This paper was prepared in collaboration with GPT-6 Astra.

\end{acknowledgments}

\section*{Data and code provenance}\label{data-and-code-provenance}

The accompanying repository preserves native Spheral snapshots,
restarts, material identities, tracer and reciprocal-gravity histories,
and the bound-companion runs. \path{result_manifest.json} and
\path{REPRODUCE.md} map results to input hashes, executed drivers,
force treatments, runtime provenance, and regeneration commands.
No public archive identifier has yet been assigned.

Particle ledgers preserve component counts, selection membership,
energy signs, and crossing flags. Exact inventories can be reproduced;
chaotic encounter histories require the stated sensitivity comparisons.
Continuation plots use saved orbital states, and analytical figures use
the stated formulas. Figure~\ref{fig:observational-context} is shared with the ablation paper.

\appendix
\FloatBarrier
\setcounter{figure}{0}\setcounter{table}{0}
\renewcommand{\thefigure}{A\arabic{figure}}\renewcommand{\thetable}{A\arabic{table}}
\renewcommand{\theHfigure}{A.\arabic{figure}}\renewcommand{\theHtable}{A.\arabic{table}}
\FloatBarrier
\section{Numerical controls and interpretation of the pilot
results}\label{appendix-a.-numerical-controls-and-interpretation-of-the-pilot-results}

The unrounded reference constants are \(G=6.67430\times10^{-11}\)
\(\mathrm{m^3\,kg^{-1}\,s^{-2}}\), \(M_T=1.34518\times10^{23}\) kg,
\(M_I=1.80566\times10^{21}\) kg, \(GM_S=3.79\times10^{16}\)
\(\mathrm{m^3\,s^{-2}}\), and \(R_S=6.0\times10^7\) m. The bound
continuations use \(J_2=0.016290573\) at \(R_{\rm ref}=60{,}330\) km.
The native inputs and saved states retain their full precision; rounding
in the main text does not change the integrations or analytical evaluations.

\begin{figure}[!htbp]
\centering
\includegraphics[width=\linewidth,keepaspectratio,alt={Oblique-run material and conservation diagnostics. The left panel distinguishes outward material mechanically unbound to the remnant from the subset that is also Saturn bound. The right panel shows the independently evaluated global energy residual with the original normalization. The classification is instantaneous and does not define a final ejecta yield.}]{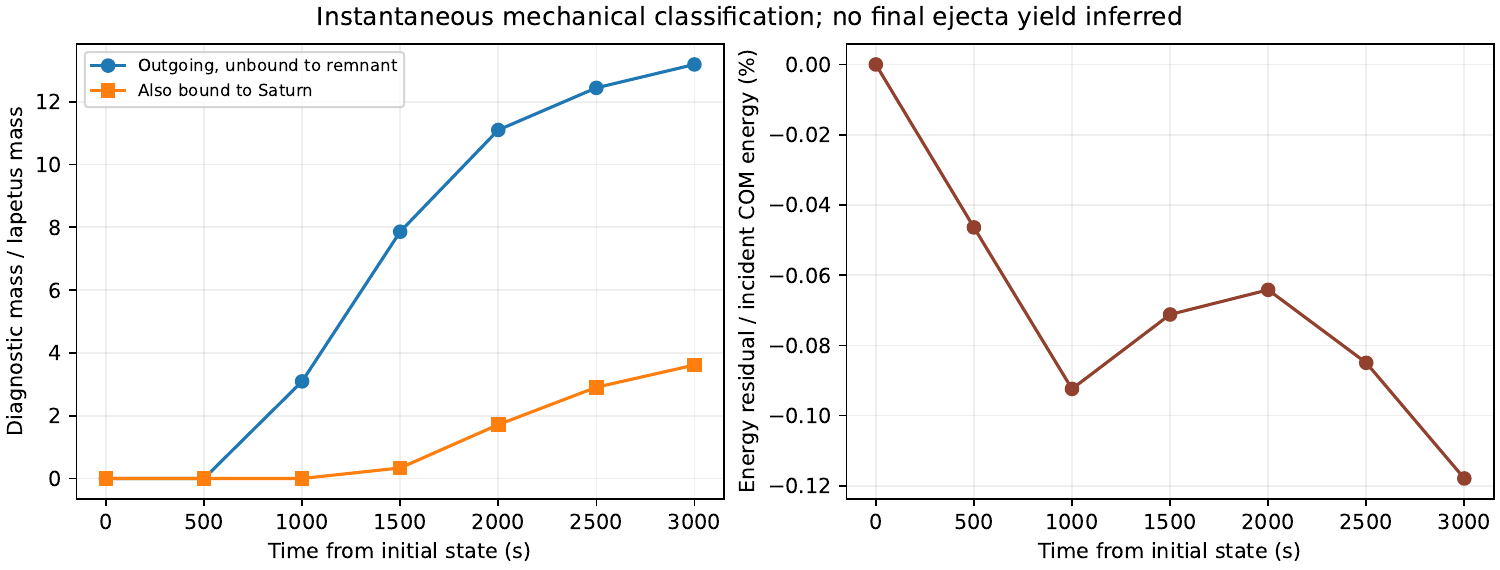}
\caption{Oblique-run material and conservation diagnostics. The left
panel distinguishes outward material mechanically unbound to the remnant
from the subset that is also Saturn bound. The right panel shows the
independently evaluated global energy residual with the original
normalization. The classification is instantaneous and does not define a
final ejecta yield.}\label{fig:diagnostics}
\end{figure}

The two 10 km s\(^{-1}\) density-summation impacts reach 3,000 s,
conserving particle mass to floating-point precision. The maximum
absolute saved energy residual across the two runs is 0.219\% of the
initial relative-motion kinetic-energy scale. The budget includes
force-consistent self-gravitational potential and Saturn's potential.
The maximum angular-momentum change is 0.0095\% of the initial
Saturn-centered total. These are global conservation diagnostics: the
large orbital angular-momentum denominator does not establish equivalent
accuracy in remnant spin, and a small global energy residual does not
establish local thermal accuracy (Figure~\ref{fig:diagnostics}).

The initial relative-motion kinetic-energy normalization \(E_{\rm rel,0}\) is

\[
E_{\rm rel,0}=\frac12\frac{M_0m_p}{M_0+m_p}u_0^2,
\tag{A1}
\]

where the actual initialized relative speed is used for a run budget.
For comparison, evaluating the same scale at the nominal 10 km
s\(^{-1}\) contact speed gives \(6.11\times10^{29}\) J. This distinction
avoids substituting contact energy for the initialized energy when
comparing precise numerical residuals.

The isolated controls produce no mechanically unbound particles during
3,000 s. However, the target's 95th-percentile particle radius decreases
by 12.2\% and the projectile's by 2.7\%; the final internal RMS
velocities are 84 and 113 m s\(^{-1}\). Their maximum energy drifts are
1.28\% and 2.34\% of the initial binding-energy magnitudes. These
controls establish that the continuous hydrostatic profiles are not yet
stationary discrete realizations. Quantitative refinement must separate
this initial adjustment from impact-driven changes.

The separate density-integration calculation stops at 704.978 s during
early release as a target-ice particle expands toward vanishing density
and its velocity-divergence timestep collapses. The native restart and
timestep vote were preserved in the run record. The density-summation
pilots restart from the initial bodies; the failed interval was not
crossed by imposing a larger timestep. This failed configuration is not
combined with the completed runs to form a continuous trajectory.

No physical fragment-size distribution is extracted from these pilots. A
hydrodynamic resolution element is neither a coherent droplet nor a
resolved gravitational aggregate by definition. The candidate masses are
sums of particle mass weights, and the material fractions refer to
immutable component labels. Likewise, the original one-way tracer
continuation omits debris reaction forces and self-gravity and does not
supply a calculation of dynamical friction. These choices set the
interpretation of the reported quantities without altering their
numerical values.

\subsection{Bound-impact energy drift and continuation
definitions}\label{a.1.-bound-impact-energy-drift-and-continuation-definitions}

The bound-impact residual is \([E(t)-E(0)]/K_{\rm rel,contact}\), where
\(E\) includes particle kinetic and thermal energy plus the exact-pair
softened self-potential and Saturn potential, and
\(K_{\rm rel,contact}=\tfrac12[M_0m_p/(M_0+m_p)]u_{\rm imp}^2\). This
contact normalization differs from equation (A1). In the \(0.10M_T\)
case it is \(8.206\times10^{28}\) J. The signed residual reaches
\(-0.57\%\) at 500 s, \(-4.57\%\) at 3,000 s, and \(-9.14\%\) at 6,000
s: half the final drift accumulates after 3,000 s. It is not confined to
a brief impact impulse.

\begin{figure}[!htbp]
\centering
\includegraphics[width=\linewidth,keepaspectratio,alt={Signed energy residuals and separated ice counts for the three bound impacts. The residual is evaluated with each run's contact relative-motion kinetic-energy normalization. The smaller companion's negative drift continues through the release interval. All three small ice inventories remain pilot yields. These diagnostics are distinct from the smaller residuals of the high-speed pilots in Figure A1.}]{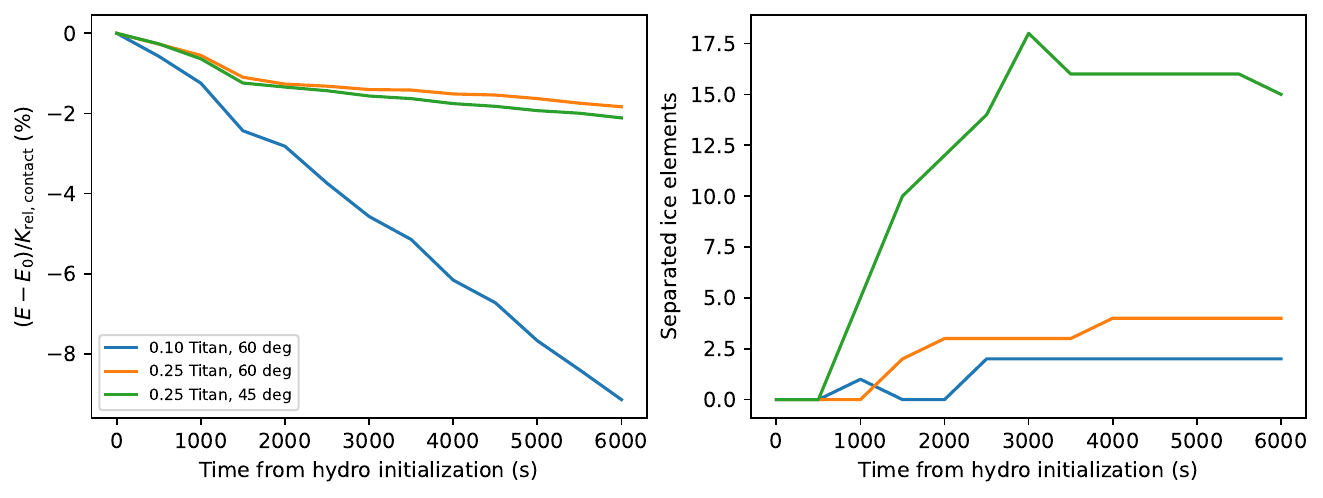}
\caption{Signed energy residuals and separated ice counts for the three
bound impacts. The residual is evaluated with each run's contact
relative-motion kinetic-energy normalization. The smaller companion's
negative drift continues through the release interval. All three small
ice inventories remain pilot yields. These diagnostics are distinct
from the smaller residuals of the high-speed pilots in Figure~\ref{fig:diagnostics}.}\label{fig:bound-energy}
\end{figure}

For the three runs in Table~\ref{tab:5}, the maximum absolute residuals
(Figure~\ref{fig:bound-energy})
are \((7.50,3.34,3.84)\times10^{27}\) J. The isolated-target
surface-binding scales \(GM_0M_i/R_T\) for their separated ice are
\((0.438,0.877,3.29)\times10^{27}\) J, giving ratios 17.1, 3.81, and
1.17. These scales are not the actual escape barriers in the interacting
system. The comparison neither assigns error to the ice nor shows it to
be spurious, but conservation alone cannot certify any of these small
yields. One element is \(0.03725M_I\), or 6.7\% of the 15-element
inventory. The saved budgets do not isolate initial adjustment,
hydrodynamic error, and tree-force versus exact-potential error;
quantitative yield calibration requires controlled refinement of
resolution and the initial state.

\begin{table}[!htbp]
\centering
\caption{Distinct numerical treatments. Each starts from its stated input; the
rows are not interchangeable stages of one uniform physical calculation.
Details and exact run identifiers are in the result manifest.\label{tab:A1}}
\small\setlength{\tabcolsep}{4pt}\renewcommand{\arraystretch}{1.15}
\begin{tabular}{@{}>{\raggedright\arraybackslash}p{(\linewidth - 4\tabcolsep)*\real{0.25000000}}>{\raggedright\arraybackslash}p{(\linewidth - 4\tabcolsep)*\real{0.27000000}}>{\raggedright\arraybackslash}p{(\linewidth - 4\tabcolsep)*\real{0.48000000}}@{}}
\toprule
Treatment & Start and duration & Evolved processes / endpoint\\
\midrule
Native Spheral impacts & Initial bodies; 3,000 or 6,000 s & Pressure, self-gravity, fixed central Saturn; material fields\\
Original one-way tracers & 3,000 s; five days & Saturn and moving-remnant forces on selected outward candidates; no
reaction or debris self-gravity\\
Native hydro/gravity overlap & 3,000 s; 300 s & Full 2,200-sample SPH versus matched pressure-free gravity\\
Reciprocal-gravity fate runs & 3,000 s; five days & One primary, all 392 unresolved samples, recoiling Saturn; no collisions
or gas\\
Bound-survivor return screens & 3,000--6,000 s; up to 50 yr & Two remnant centers, all other samples, reciprocal gravity and Saturn
\(J_2\); approach or time limit\\
Direct compact contact & Saved \(45^\circ\) approaches; 22.8 or 29.7 min & Same gravity to first contact, stopping before hydrodynamics\\
Complete-merger estimate & Saved contact state & Algebraic mass and momentum sum with zero new ejecta\\
\bottomrule
\end{tabular}
\end{table}

The reference 5,000-s \(45^\circ\) screen uses IAS15 tolerance
\(10^{-10}\), a 600-s trial-step cap, and a 5,555-km screening radius
sum. Its direct continuation stops at the compact 4,437-km sum. The
tighter comparison uses \(10^{-11}\), a 300-s cap, and compact screening
throughout. Their first screened flyby has predicted pericenter 7,083.60
versus 7,083.29 km. Common hourly states first exceed a 1-km
relative-position difference near day 612.67, 100 km near day 625.29,
and 1,000 km near day 768.25. This locates the divergence well before
the illustrative return; hourly sampling does not independently resolve
all intervening encounters. These comparisons concern the 5,000-s
handoff, separately from the controlled later-handoff restarts below.

The four 6,000-s continuations in Table~\ref{tab:7} resume the same saved
30-yr state, with 18 mutually gravitating bodies, unchanged softening,
Saturn \(J_2\), and a compact contact-radius sum of 4,443 km. Only the
step cap and IAS15 tolerance change. The restarted position, velocity,
and mass arrays match the parent exactly; a four-day continuous run and
two successive two-day restarts agree at every common hourly state.
Recorded relative energy residuals remain below \(1.5\times10^{-15}\)
in these extensions, but do not establish encounter convergence. The
original continuation reaches its approach screen before direct
integration to contact; the other three stop at 50 yr. Only the axial
angular-momentum component is conserved by the fixed-axis \(J_2\)
potential. No second hydrodynamic collision is calculated.

\FloatBarrier
\setcounter{figure}{0}\setcounter{table}{0}
\renewcommand{\thefigure}{B\arabic{figure}}\renewcommand{\thetable}{B\arabic{table}}
\renewcommand{\theHfigure}{B.\arabic{figure}}\renewcommand{\theHtable}{B.\arabic{table}}
\FloatBarrier
\section{Analytical checks and source
distinctions}\label{appendix-b.-analytical-checks-and-source-distinctions}

\path{analytical_checks.py} reproduces the impulse scale, the
historical ejecta and useful-fraction benchmarks, particle mass
resolution, growth and scattering scales, the fixed-\(h\) bounds, and
the local gas-drag estimate. The archived particle tally supplies a
common numerical source for Table~\ref{tab:2} and Figure~\ref{fig:partition}. Figure~\ref{fig:angular-momentum} is generated
from the analytical bound; Figure~\ref{fig:impact} and Figure~\ref{fig:diagnostics} retain the existing
pilot visualizations. No time-dependent particle data are reconstructed
from plotted pixels.

\path{bound_satellite/titan_flyby_impulse.py} reproduces equation (27)
and the encounter scales in Section~\ref{close-encounters-and-circularization}.
It uses the saved 30-yr icy orbit in an explicitly constructed crossing
geometry, includes Titan's recoil, neglects Saturn's differential force
during the short encounter, and recomputes eccentricity from the outgoing
velocity. Let \(b_{\infty,\max}\) and \(b_{\infty,\mathrm{contact}}\)
be the asymptotic impact parameters for the largest separation meeting
the eccentricity-decrease criterion and for surface contact. The area bound uses
\(b_{\infty,\max}^2/b_{\infty,\mathrm{contact}}^2-1\); it counts all
orientations before selecting those that lower eccentricity. The archived
checks verify the kick formula, relative speed, two-body momentum, and
scalar/vector eccentricity agreement.

The observational mass and orbit scales added to the contextual
discussion are from \citet{jpl2026a,jpl2026b}. The
debris-damping and gas-response literature supplies physical comparisons
in Section~\ref{can-the-debris-circularize-iapetus-or-is-gas-required}, not replacement simulation results. The donor-mass example
in equation (20) and the local drag estimate in equation (25) are
additional analytical scale evaluations. They are not measurements of
the pilot debris or an adopted global residual disk. The comparison with
the original 2005 scenario retains its stated ejecta, growth,
scattering, and gas-column scales.

The distinction between a final semimajor axis and an apocenter is
checked explicitly in the supplement: \((q,Q)=(20,60)R_S\) gives
\((a,e,a_{\rm circ})=(40R_S,0.5,30R_S)\), whereas \((a,e)=(60R_S,0.7)\)
gives \((q,Q,a_{\rm circ})=(18,102,30.6)R_S\). These examples are two
different post-scattering orbits. No orbital probability is attached to
either.

\path{asphaug_comparison_checks.py} evaluates the normalized impact
speed, nominal contact angles, and the target-ice lower bound in
equations (12a), (B1), and (B2) from the recorded inputs and the
unrounded particle tally underlying Table~\ref{tab:2}. It also records the minimum
projectile-rock mass outside both the dominant remnant and the selected
subset. These are conservation and scaling deductions, not additional
impacts or reconstructed particle histories. The comparison to \citet{asphaug2006} uses the parameters and description of their
Figure 3b. Equation (26) states the momentum accounting for a return
collision if one occurs; it is not an additional simulated event. The
final Titan eccentricity must be obtained from the fully evolved remnant
state, not inferred from the first-impact material table.

The tagged target-ice measurement supersedes the older lower-bound
argument as the primary evidence. As an independent conservation check,
the initial projectile ice mass \(M_{i,p,0}\) is

\[
M_{i,p,0}=0.30(0.1M_T)=2.235M_I,\tag{B1}
\]

whereas the selected oblique ice mass \(M_{i,\mathrm{selected}}\) is
\(2.9054M_I\). Even if all projectile ice entered the selection, its
target-derived contribution \(M_{i,T,\mathrm{selected}}\) would satisfy

\[
M_{i,T,\mathrm{selected}}\geq M_{i,\mathrm{selected}}-M_{i,p,0}\simeq0.670M_I.\tag{B2}
\]

The measured target contribution, \(2.7564M_I\), is much larger.

The original one-way tracer continuation starts with all remnant-unbound
outward candidates, including positive Saturn-energy material. In the
oblique case all 97 initially negative-energy tracers remain
negative-energy after five days, and 21 others change sign: five
target-ice, two projectile-ice, and 14 projectile-rock tracers. Thus
\(3.6132+0.7822=4.3954M_I\), with final ice fraction 0.7203. In the
head-on case, 49 of the initial 57 remain bound and eight intersect
Saturn: \(2.1232-0.2980=1.8252M_I\). No additional initially
positive-energy tracer becomes bound. These exactly closed transitions
describe the original tracer force model, not the later full-inventory
reciprocal-gravity calculation. Five days is about 0.32 Titan or 0.062
Iapetus orbital periods at the reference radii.

\FloatBarrier
\setcounter{figure}{0}\setcounter{table}{0}
\renewcommand{\thefigure}{C\arabic{figure}}\renewcommand{\thetable}{C\arabic{table}}
\renewcommand{\theHfigure}{C.\arabic{figure}}\renewcommand{\theHtable}{C.\arabic{table}}
\FloatBarrier
\section{Numerical debris controls and physical-fragment
assumptions}\label{appendix-c.-numerical-debris-controls-and-physical-fragment-assumptions}

The compact-group finder uses spatial linking and iterative mechanical
unbinding. Each particle's reference spacing is the cube root of its mass
divided by its component's reference density. A pair is linked when its
separation is below the linking factor times the mean of their two
spacings. At 1.5 reference particle spacings it gives a 1,808-element
primary, an 18-element secondary candidate (16 projectile rock and two
projectile ice), and 374 unresolved samples. The secondary has
\(0.670M_I\) and 88.9\% rock. At links 1.25 and 1.75, a secondary above
the 16-element reporting threshold is respectively absent or
\(1.676M_I\). The primary-only handoff leaves all other 392 samples
distributed. The threshold does not decide whether a smaller persistent
concentration exists.

\begin{table}[!htbp]
\centering
\caption{Fixed-material hydrodynamic diagnostics. Cohorts are nested sets,
mechanically self-bound in their own COM frame with the native softened
potential and no thermal term. RMS radii are particle-center extents,
not physical surfaces.\label{tab:C1}}
\small\setlength{\tabcolsep}{4pt}\renewcommand{\arraystretch}{1.15}
\begin{tabular}{@{}>{\raggedright\arraybackslash}p{(\linewidth - 8\tabcolsep)*\real{0.18000000}}>{\raggedright\arraybackslash}p{(\linewidth - 8\tabcolsep)*\real{0.20500000}}>{\raggedright\arraybackslash}p{(\linewidth - 8\tabcolsep)*\real{0.20500000}}>{\raggedright\arraybackslash}p{(\linewidth - 8\tabcolsep)*\real{0.20500000}}>{\raggedright\arraybackslash}p{(\linewidth - 8\tabcolsep)*\real{0.20500000}}@{}}
\toprule
Rock cohort & Time (s) & RMS radius (km) & Internal RMS speed (m s\(^{-1}\)) & Saturn energy (MJ kg\(^{-1}\))\\
\midrule
Inner (16) & 3,000 & 478.2 & 212.9 & +0.902\\
Inner (16) & 3,300 & 483.5 & 209.9 & +0.886\\
Wider (43) & 3,000 & 949.3 & 277.7 & +1.730\\
Wider (43) & 3,300 & 986.7 & 273.3 & +1.659\\
\bottomrule
\end{tabular}
\end{table}

At the tight link, 14 inner-cohort elements belong to a 15-element
group; two are in a pair and a singleton. The combined cohort
nevertheless remains self-bound. Its final 300-s RMS-radius growth is
1.12\%. At 3,300 s the median smoothing scale is 2,965 km, about 6.1
times its RMS radius: the fluid structure is poorly resolved. The
cohorts initially separate from Titan at 4.4--4.5 km s\(^{-1}\) while
moving inward around Saturn at about 4.5 km s\(^{-1}\). Their
positive-energy states motivate following escape without imposing a
return.

The native 3,000--3,300 s SPH overlap and a force-matched gravity-only
calculation of all 2,200 samples differ by 6.20 m s\(^{-1}\) RMS outside
the two compact candidates. The 95th-percentile velocity difference
normalized by initial remnant-relative speed is 0.100\%, with a 100 m
s\(^{-1}\) floor. This supports a short orbital handoff without
validating pressure-free evolution for five days. SHARD provides a
precedent for numerical debris reduction \citep{crespi2026}; the controls below test our own mass-weighted
phase-space subdivision.

The compression controls start from the full oblique snapshot and retain
every target/projectile rock and water sample. For the two-candidate
comparison, 374 unresolved samples contain \(13.9311M_I\); Saturn and
the two compact candidates bring the gravitating count to 377. Retaining
the 18 candidate-core particles separately instead gives 392 unresolved
samples and 394 gravitating objects including Saturn. The two policies
are alternative numerical treatments of the same material inventory, not
different source populations. In the two-candidate realization, 33
debris samples totaling \(1.2292M_I\) have negative instantaneous energy
in the softened Titan-only potential. They remain debris rather than
being immediately merged into Titan.

For compression, positions and velocities are scaled by 1,000 km and 1
km s\(^{-1}\). The cell with the largest mass-weighted squared
dispersion is divided along its most dispersed coordinate at a mass
median until the requested count is reached. Parcel masses are generally
unequal. Every cell records its members, component masses,
center-of-mass state, full position-velocity covariance, thermal energy,
unresolved kinetic energy, and internal angular momentum. The maximum
normalized residual across the mass, first-moment, momentum,
kinetic-energy, and angular-momentum identities is
\(4.83\times10^{-13}\), including the collision-COM frame. This closes
the ledger but does not evolve the recorded internal moments.

All pairs exert reciprocal gravity; there is no active/test-particle
split. For source mass \(m\), pair separation \(r\), and softening
length \(\epsilon\), non-Saturn pairs use the Spheral pilot's softened
radial acceleration, with magnitude \(Gm/(r^2+\epsilon^2)\) and
\(\epsilon=136.502\) km. Its specific potential is \(-Gm\arctan(\epsilon/r)/\epsilon\). Saturn's 100-km
Plummer potential is also made reciprocal. Softening is a numerical
force parameter, not a physical fragment radius. The reference IAS15
tolerance is \(10^{-10}\) with a 30-s trial-step cap. Each control lasts
300 s, with collisions disabled and no assigned parcel collision radii.

\begin{figure}[!htbp]
\centering
\includegraphics[width=\linewidth,keepaspectratio,alt={Numerical compression of the oblique debris sample, using the same provisional two-remnant selection. Left: initial velocity dispersion recorded but absent from centroid motion. Right: parcel-centroid velocity error after 300 s relative to the full 374-sample calculation. The velocity-only control uses the same mass-weighted subdivision algorithm with position omitted. These are collisionless representation tests, not fragment-size or satellite-formation predictions.}]{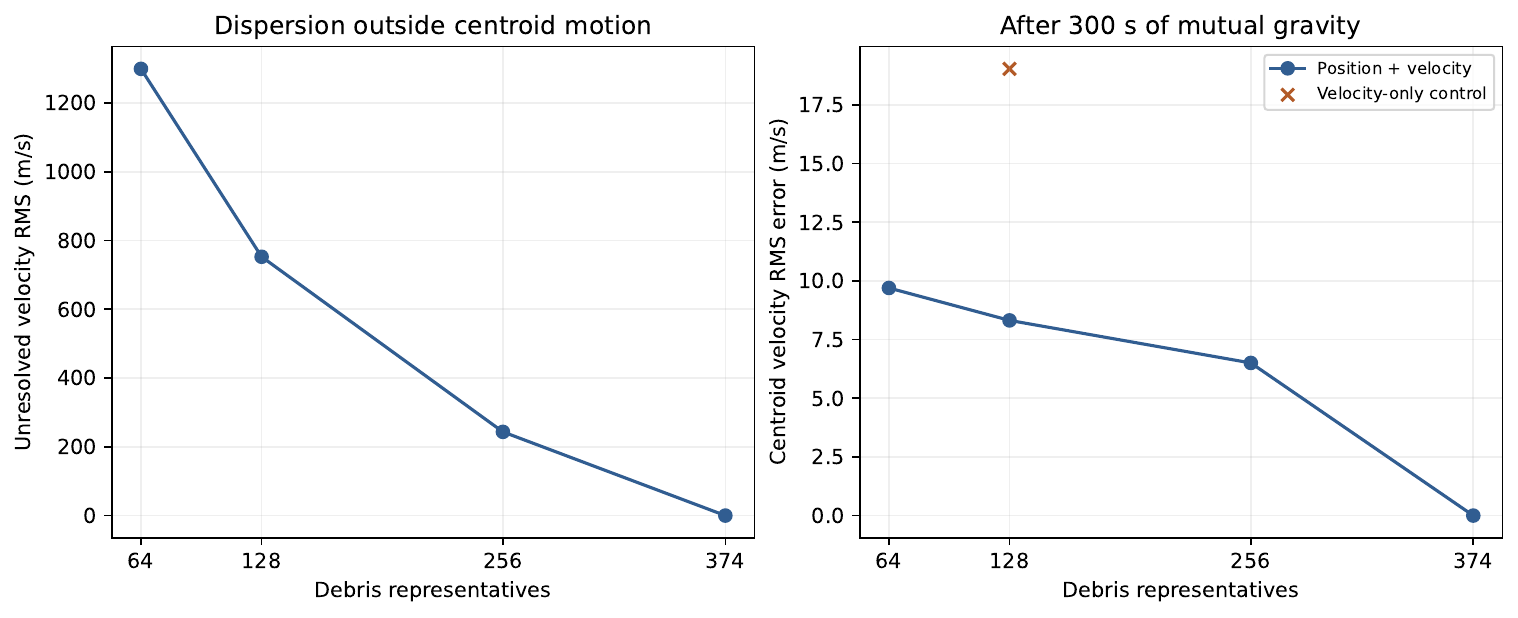}
\caption{Numerical compression of the oblique debris sample, using the
same provisional two-remnant selection. Left: initial velocity
dispersion recorded but absent from centroid motion. Right:
parcel-centroid velocity error after 300 s relative to the full
374-sample calculation. The velocity-only control uses the same
mass-weighted subdivision algorithm with position omitted. These are
collisionless representation tests, not fragment-size or
satellite-formation predictions.}\label{fig:compression}
\end{figure}

Using 64, 128, and 256 representatives leaves initial unresolved
velocity dispersions of 1,299, 753, and 243 m s\(^{-1}\) RMS. After 300
s, the corresponding parcel-centroid velocity errors relative to the
moving centroids of the full 374-sample reference are 9.70, 8.32, and
6.51 m s\(^{-1}\) RMS. These latter errors exclude the internal spread,
rather than proving that it is dynamically unimportant. A velocity-only
subdivision with 128 representatives gives 19.04 m s\(^{-1}\); changing
the spatial scale to 3,000 km gives 14.37 m s\(^{-1}\). Neither
comparison reproduces SHARD's k-means algorithm. Figure~\ref{fig:compression} summarizes
the count comparison.

Separate controls quantify changes in the force representation.
Collapsing both candidate remnants changes debris velocities by 5.27 m
s\(^{-1}\) RMS relative to a 2,200-element, gravity-only reference; that
reference itself omits pressure support. Doubling softening changes the
full 374-sample debris velocities by 7.08 m s\(^{-1}\) RMS. Tightening
the integration tolerance to \(10^{-11}\) and halving the step cap
changes the 128-parcel centroid velocities by less than
\(2\times10^{-12}\) m s\(^{-1}\) RMS. Across the ten baseline controls,
the maximum saved relative energy and orbital-angular-momentum errors
are \(2.09\times10^{-15}\) and \(1.78\times10^{-15}\). These distinguish
integration accuracy from physical fidelity and motivate retaining the
full sample here.

Earlier exploratory runs instead assumed instantaneous condensation into
individual physical fragments of 0.01862 or 0.00931 Iapetus masses,
assigning collision radii from those masses and material densities.
Their assumed merger, rebound, and fragmentation prescriptions reached
unresolved collision geometries after 1,056 and 376 s, respectively.
Those stops concern that physical-fragment prescription. They neither
establish the fate of a representative debris distribution nor determine
a fragment-size spectrum. The statistical collision extension requires
physical constituent masses and sizes independently of numerical parcel
weights, with consistent exchange of energy and angular momentum. It has
not been implemented in the representative controls. No completed gas
comparison or automated Spheral collision-return calculation is
reported.

\subsection{Forces, residuals, and event
definitions}\label{c.1.-forces-residuals-and-event-definitions}

Let \(E_{\rm mech}\) be the total mechanical energy: translational
kinetic energy, all reciprocal pair potentials, and the Saturn potential.
The gravity-only residual is
\([E_{\rm mech}(t)-E_{\rm mech}(0)]/|E_{\rm mech}(0)|\). Archived thermal energy, internal dispersion, and spin
neither enter this denominator nor feed back dynamically. A small
full-system residual therefore cannot certify debris/core trajectory accuracy. Let
\(\boldsymbol L\) be the total orbital angular momentum and
\(L_z\) its component along Saturn's fixed symmetry axis. The unbound
runs use \(|\boldsymbol L(t)-\boldsymbol L(0)|/|\boldsymbol L(0)|\); the
axisymmetric bound runs test \(|L_z(t)-L_z(0)|/|L_z(0)|\). Their
separately recorded full-vector changes can include physical precession.

Saturn's fixed \(J_2\) axis is the initial \(z\) axis, perpendicular to
the impact plane. Its force acts reciprocally between Saturn and every other gravitating object. For object masses \(m_i\),
Saturn-centered distances \(r_i\), and axial coordinates \(z_i\), the
potential-energy contribution is
\(GM_SJ_2R_{\rm ref}^2\sum_i m_i(3z_i^2/r_i^2-1)/(2r_i^3)\), in addition
to the Plummer monopole. The orbital energy diagnostic includes this
term. Solar perturbations and higher harmonics are omitted.

The five-day geometric test flags crossings when the minimum
straight-segment distance between accepted integrator endpoints falls below
\(R_S\) or the compact primary's initial 95th-percentile particle radius
\(R_{95}\). Bound-return debris uses \(R_S\) and the larger of the
primary's measured \(R_{95}\) and component-volume radius, also without
removal. These screens are distinct from remnant contact. With component
masses \(M_j\) and densities \(\rho_j\), compact radii \(R\) use \(R=[3\sum_jM_j/\rho_j/(4\pi)]^{1/3}\) with rock and ice densities
2,900 and 917 kg m\(^{-3}\). A softened two-body pericenter estimate determines whether to stop a returning approach.
For companion and primary centers \(\boldsymbol x_c,\boldsymbol x_T\)
and adopted radii \(R_c,R_T\), direct contact evolves the same force until
the surface gap
\(g(t)=|\boldsymbol x_c-\boldsymbol x_T|-(R_c+R_T)\) reaches 1 m,
limiting trial steps to 1 s and one quarter of gap divided by relative
speed. Only the separate algebraic estimate assumes merger.

\clearpage
\bibliographystyle{aasjournalv7.1}
\bibliography{references}

\end{document}